\documentclass{svjour2}                    
\smartqed  
\usepackage{graphicx}
\usepackage{amssymb}
\def\H{H\hskip-8.5pt/\hskip2pt}
\def\e{{\rm e}}
\def\coeff#1#2{{\textstyle{#1\over #2}}}

\def\lsim{\mathrel{\mathpalette\@versim<}}

\def\gsim{\mathrel{\mathpalette\@versim>}}

\begin{document}

\title{Decoherence and CPT Violation in a Stringy Model of Space-Time Foam
}


\author{Nick E. Mavromatos
}


\institute{Nick E. Mavromatos \at
              King's College London, Department of Physics, Strand, London WC2R 2LS, U.K. \\
              \email{Nikolaos.Mavromatos@kcl.ac.uk}           
}

\date{Received: date / Accepted: date}

\maketitle

\begin{abstract}
I discuss a model inspired from  the string/brane framework, in which our Universe is represented (after perhaps appropriate compactification) as a three brane, propagating in a bulk space time punctured by D0-brane (D-particle) defects. As the D3-brane world moves in the bulk, the D-particles cross it, and from an effective observer on D3 the situation looks like a ``space-time foam'' with the defects ``flashing'' on and off (``D-particle foam''). The open strings, with their ends attached on the brane, which represent matter in this scenario, can interact with the D-particles on the D3-brane universe in a topologically non-trivial manner, involving splitting and capture of the strings by the D0-brane defects.
Such processes are consistently described by logarithmic conformal field theories on the world-sheet of the strings. Physically, they result in effective decoherence of the string matter on the D3 brane, and as a result, of CPT Violation, but of a type that implies an ill-defined nature of the effective CPT operator. Due to electric charge conservation, only electrically neutral (string) matter can exhibit such interactions with the D-particle foam. This may have unique, experimentally detectable (in principle), consequences for electrically-neutral entangled quantum matter states on the brane world, in particular the modification of the pertinent Einstein-Podolsky-Rosen (EPR) Correlation in neutral mesons in an appropriate meson factory. For the simplest scenarios, the order of magnitude of such effects might lie within the sensitivity of upgraded $\phi$-meson factories.

\end{abstract}

\section{Introduction}
\label{intro}

In the recent decade, several authors~\cite{kostelecky,lorentz} in the physics community, in an attempt to discuss quantum gravity scenarios, have considered speculative theoretical models which violate or are characterised by modified Lorentz symmetry in space time and, as a result, their quantum versions (wherever they exist) \emph{may} exhibit CPT-symmetry breakdown~\footnote{For some of the modified Lorentz Symmetry models, the
situation concerning CPT symmetry is unclear, given that their quantum versions are not understood fully as yet.}. CPT invariance is guaranteed by a mathematical theorem~\cite{cpt} applicable to flat-space-time relativistic quantum  field theories, which respect unitarity, locality and Lorentz symmetry. As discussed in ref.~\cite{greenberg}, Lorentz invariance is a fundamental reason to guarantee CPT conservation. If CPT is violated, then Lorentz symmetry must have been violated, but not vice versa, in other words one might have terms in an effective local field theory, for instance the so-called Standard Model Extension~\cite{kostelecky}, that violate Lorentz symmetry but \emph{not} CPT. But the CPT Violating terms in the Lagrangian are necessarily Lorentz Violating.

All the above-mentioned features are based on the local effective lagrangian formalism, with the Lorentz- and/or CPT -Violating  interactions being represented by appropriate local (but higher-dimensional, non renormalizable) operators, suppressed by appropriate negative powers of the effective mass scale that characterises the relevant set of degrees of freedom. We should remark at this stage that this may not be
necessarily that  of quantum gravity $M_{\rm QG} \sim 10^{19}$~GeV, although in many theories is. For instance, in models of quantum electrodynamics~\cite{hathrell} in a \emph{classical curved background}, the integration of electron loops gave terms in the effective lagrangian, coupling the Maxwell field-strength tensor to appropriate curvature tensors, which have been suppressed by the mass of the electron that played the r\^ole of the effective scale in this case. Of course, such models did not consider quantum fluctuations of the gravitational fields, but nevertheless the presence of curvature violated flat-space Lorentz invariance. This is the reason why I brought this example up. Since we do not understand microscopically as yet the reason for possible Lorentz Violation in an effective low-energy field theory, we should keep an open mind to all possibilities.
In fact, when one considers extensions of the standard model~\cite{kostelecky}, it is tempting to make naive dimensional analysis estimates of the various dimensionful coefficients of the Lorentz-symmetry-violating terms by having the Planck Mass as a reference scale. In view of the above-mentioned example of macroscopic curvature coupling to the photon field-strength terms, this may be misleading.

A true theory of quantum gravity may or may not violate such symmetries.
However it may also imply structures beyond the above-mentioned local effective lagrangians. The example of an evaporating black hole or that of a collapsing matter to form a black hole are two typical examples, where the relevant processes cannot be described in terms of local operators in an effective field theory with a well defined scattering matrix, or at least in the way we formulate the scattering matrix today, by means of asymptotic states. Indeed, the presence of horizons complicate matters. As suggested by Hawking, one might have degrees of freedom crossing such horizons which are lost from a low-energy observer in a quantum gravity setting~\cite{hawking}, where microscopic (\emph{i.e.} of the size of Planck length $1/M_{\rm QG} $) horizons are present in a path integral over quantum fluctuations in space time.
In fact, the structure of space-time at Planck scales may even be discrete and topologically non-trivial, of a  ``foamy'' nature, so to speak~\cite{wheeler}.
In this sense, the matter system, after integration over quantum gravitational  degrees of freedom,
which a low-energy local observer has no access to, becomes an \emph{open} quantum system, entailing decoherence of quantum matter, and thus evolution of initially pure states to mixed ones. The set of (integrated out) quantum-gravity degrees of freedom plays the r\^ole of the ``environment''.
We should stress at this point that, in our opinion, such a ``loss of information'' in a theory of quantum gravity is only \emph{apparent}, in the sense that it pertains to matter (low-energy) local observers, who can measure things only by means of scattering. The full theory of quantum gravity
is hopefully consistent and unitary, but unfortunately such a theory at present is not understood.

Recently these quantum-gravity-induced decoherence ideas have been rejected by their  proposer~\cite{hawking2}. Hawking has argued, based on earlier proposal by Maldacena
for a discussion of black hole evaporation in the context of strings~\cite{malda}, that
in a Euclidean path integral formalism -- which according to Hawking is the only mathematically consistent way to perform a quantum gravity path integral over both geometries and topologies of space-time -- the contributions from the topologically non-trivial configurations, which would be responsible for loss of information by a low-energy observer, decay exponentially in time, thereby leaving only the trivial-topology, unitary, contributions. This approach was achieved by regularising the space time by a small negative cosmological constant (thus making it anti-de-Sitter), which is known to have holographic properties~\cite{malda2}, and then removing the cosmological-constant regulator.

We must say that we find the situation far from being resolved.
The arguments against decoherence presented in \cite{hawking2} pertain to a specific model for decoherence, that of an evaporating black hole, and actually of a particular kind, in an anti-de-Sitter type space time.
First it is not clear to us whether this exhaust all the sectors in a full theory of quantum gravity, and hence we are unsure that the path integration is done fully (not to mention its Euclidean nature by construction, which is another major problematic aspect of quantum gravity path integration).
Second, as we shall demonstrate below, the evaporating black hole  case is not the only case where decoherence of matter might occur. Indeed, as we shall discuss in this article, and have analysed in our works in the past~\cite{Dfoam}, there are models, notably some in string theory with defects in space time (branes), that involve processes leading to decoherence of matter, in the sense of information flow to degrees of freedom pertinent to fluctuations of space time that are not detected by local scattering experiments.
In this sense, from the point of view of low-energy observers, performing (scattering) experiments, the matter system appears as decohering.

In fact, as we shall discuss below, the system violates locally Lorentz symmetry, which however is preserved globally, in the sense of the relevant symmetry-violating observables having zero expectation values. However, Lorentz-violations can occur in the fluctuations of these observables, which are non zero and lead to decoherence of the quantum matter system. In this sense one is lead to a breakdown of CPT Symmetry.
However, the type of violation is different from the one discussed in \cite{greenberg,kostelecky,lorentz} in the sense that decoherence is a process that cannot be described in the context of local effective lagrangians by means of non-renormalisable higher-dimension local field-theory operators in an effectively flat space time.
The evolution of pure to mixed states implies an \emph{irreversible} process, a \emph{microscopic time arrow}, which according to a theorem by R. Wald~\cite{wald} implies that the fundamental quantum-mechanical operator responsible for the generation of the CPT transformation, is \emph{ill-defined} in the subspace of only-matter degrees of freedom.
In this sense, one has an \emph{apparent intrinsic violation} of the CPT symmetry, which is qualitatively different from the mere non-commutativity of the (well-defined) CPT generator with the effective local Hamiltonian density of the system in the local (Lorentz and/or CPT -violating) effective-field-theory  approach to quantum gravity of refs.~\cite{kostelecky,lorentz}. In the decoherence case, as already mentioned, the CPT operator is not well defined. Due to the weakness of quantum gravitational interactions, however, this ill-defined nature is \emph{perturbative}, in the sense that the anti-particle state still exists, but it has slightly modified properties from the cases of the local effective lagrangians, where CPT (even if non commuting with the Hamiltonian) is well defined as a quantum mechanical operator.

As discussed in \cite{bmp,bmpw}, and shall review here,  this
ill-defined nature of CPT,
induced by quantum-gravity decoherence, has important phenomenological consequences in entangled states of matter, associated with the modification of the pertinent Einstein-Podolsky-Rosen (EPR) correlations, which are rather unique to this type of CPT violation. In fact, for the particular type of D-particle foam, which we concentrate our attention upon in this work, the maximum possible order of magnitude of such effects can~\cite{bernabeu} lie within the sensitivity of future neutral-Kaon facilities in an upgrade of the DA$\Phi$NE detector~\cite{dafne}.

The structure of the article is the following: in the next section \ref{sec:2} we shall discuss the mathematical aspects of D-particle foam and trace the origin of quantum decoherence of low-energy matter propagating in the foam to the loss of conformal invariance on the world-sheet, as a result of topologically non-trivial interactions of matter with the D-particle defects (capture and splitting of open matter strings by D-particles).
In section \ref{sec:3} we shall apply this formalism to discuss issues of CPT breakdown, in the sense of a CPT generator with an ill-defined nature due to decoherence, in an entangled system of neutral
bosons, representing, say neutral kaons in a $\phi$ factory. As we shall see, the order of magnitude of the associated EPR modifications, due to the perturbatively ill-defined-nature of the CPT operator, can be falsifiable ---at least in the most naive of models of D-particle foam -- in the next generation $\phi$-factory facilities.
Conclusions and outlook, regarding examining cosmological properties of D-particle foam,
and hence constraining it further (and more stringently) by means of astrophysical observations, are presented in section~\ref{sec:4}.

\section{D-particle Foam and Quantum Decoherence in String Theory \label{sec:2}}

\subsection{General Formalism of D-particle/String Interactions\label{subsec:lcft}}

\begin{figure}[t]
\centering
\includegraphics[width=7.5cm]{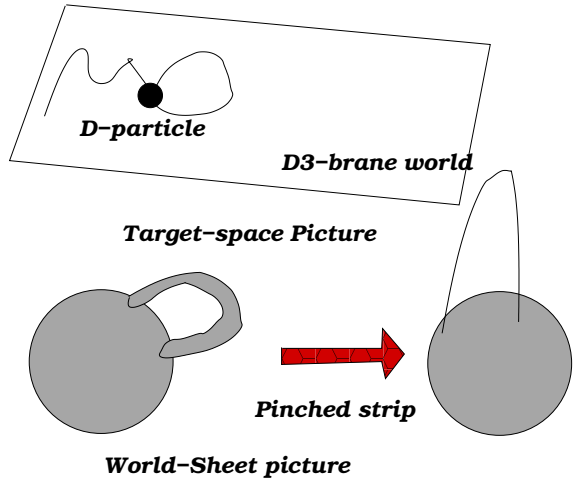}\caption{\emph{Upper picture:} A
fluctuating D-particle is described by open strings attached to it. As a
result of conservation of  string fluxes~\cite{polch,polch2,johnson} that accompany
the D-branes, an isolated D-particle cannot occur, but it has to be connected
to a D-brane world through flux strings. \emph{Lower picture}: World-sheet
diagrams with annulus topologies, describing the fluctuations of D-particles
as a result of the open string states ending on them. Conformal invariance
implies that pinched surfaces, with infinitely long thin strips, have to be
taken into account. In bosonic string theory, such surfaces can be
resummed~\cite{szabo}. }%
\label{fig:dbranes}%
\end{figure}

In this section we shall review briefly the basic features of the D-particle
foam model, discussed in \cite{Dfoam}.
We will use some established results and constructs from string/brane theory
\cite{polch,polch2}, which we shall discuss briefly for the benefit of the non-expert reader.
In particular, zero dimensional D-branes \cite{johnson}
occur (in bosonic and some supersymmetric string theories) and are also known
as D-particles. Interactions in string theory are, as yet, not treated as
systematically as in ordinary quantum field theory where a second quantised
formalism is defined. The latter leads to the standard formulations by
Schwinger and Feynman of perturbation series. When we consider stringy matter
interacting with other matter or D-particles, the world lines traced out by
point particles are replaced by two-dimensional world sheets. World sheets are
the parameter space of the first quantised operators ( fermionic or bosonic)
representing strings. In this way the first quantised string is represented by
actually a two dimensional (world-sheet) quantum field theory. An important
consistency requirement of this first quantised string theory is conformal
invariance which determines the space-time dimension and/or structure. \ This
symmetry permits the representation of interactions through the construction
of measures on inequivalent Riemann surfaces \cite{green}. In and out states
of stringy matter are represented by vertex insertions at the boundaries. The
D-particles as solitonic states~\cite{polch2} in string theory do fluctuate
themselves quantum mechanically;  this is described by stringy excitations, corresponding to
open strings with their ends attached to the D-particles and higher
dimensional D branes. In a first quantised (world-sheet) language, such
fluctuations are also described by Riemann surfaces of higher topology with
appropriate Dirichlet boundary conditions (c.f. fig.~\ref{fig:dbranes}). The
plethora of Feynman diagrams in higher order quantum field theory is replaced
by a small set of world sheet diagrams classified by moduli which need to be
summed or integrated over \cite{zwiebach}. 

\begin{figure}[th]
\centering
\includegraphics[width=7.5cm]{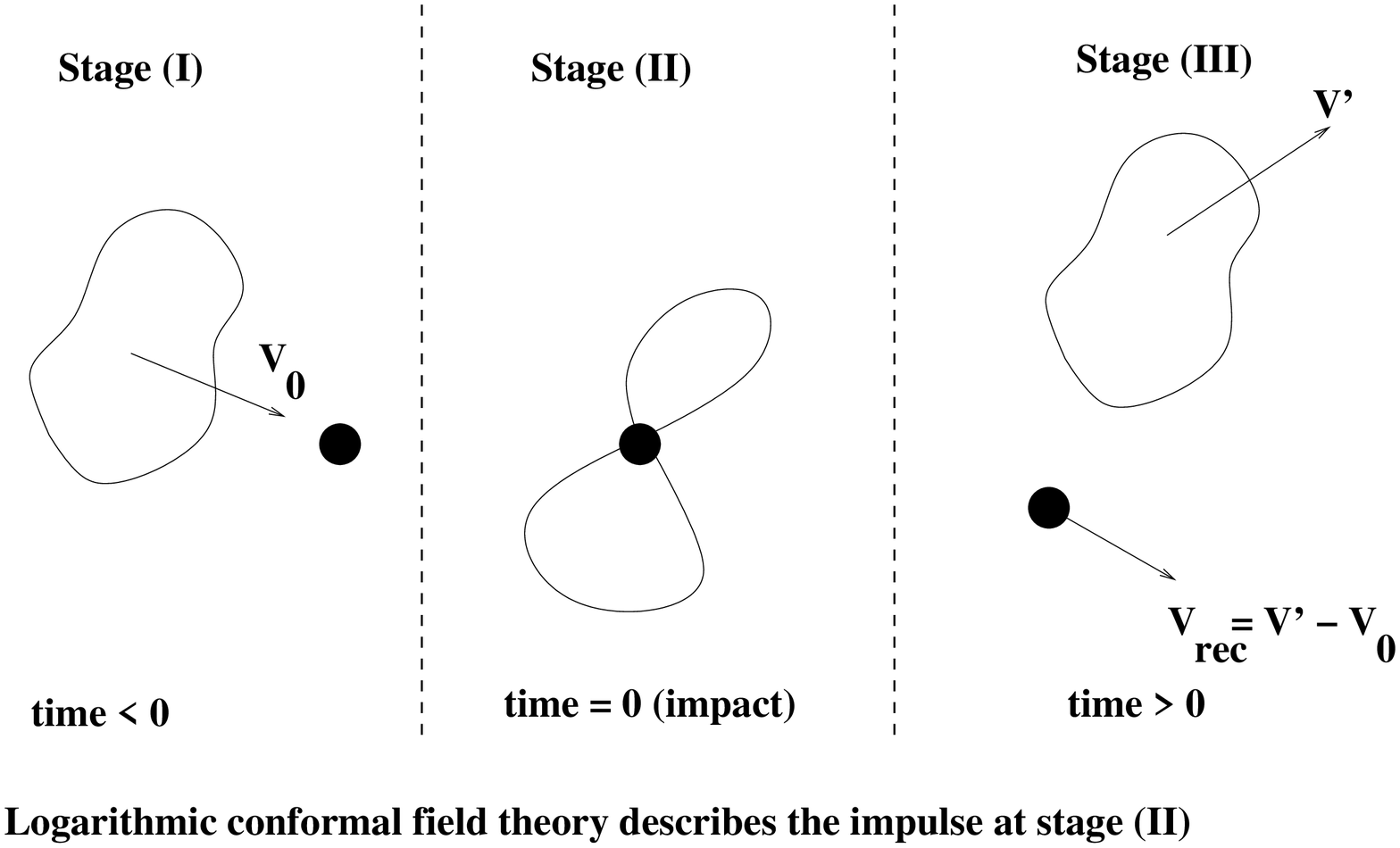} \hfill
\includegraphics[width=7.5cm]{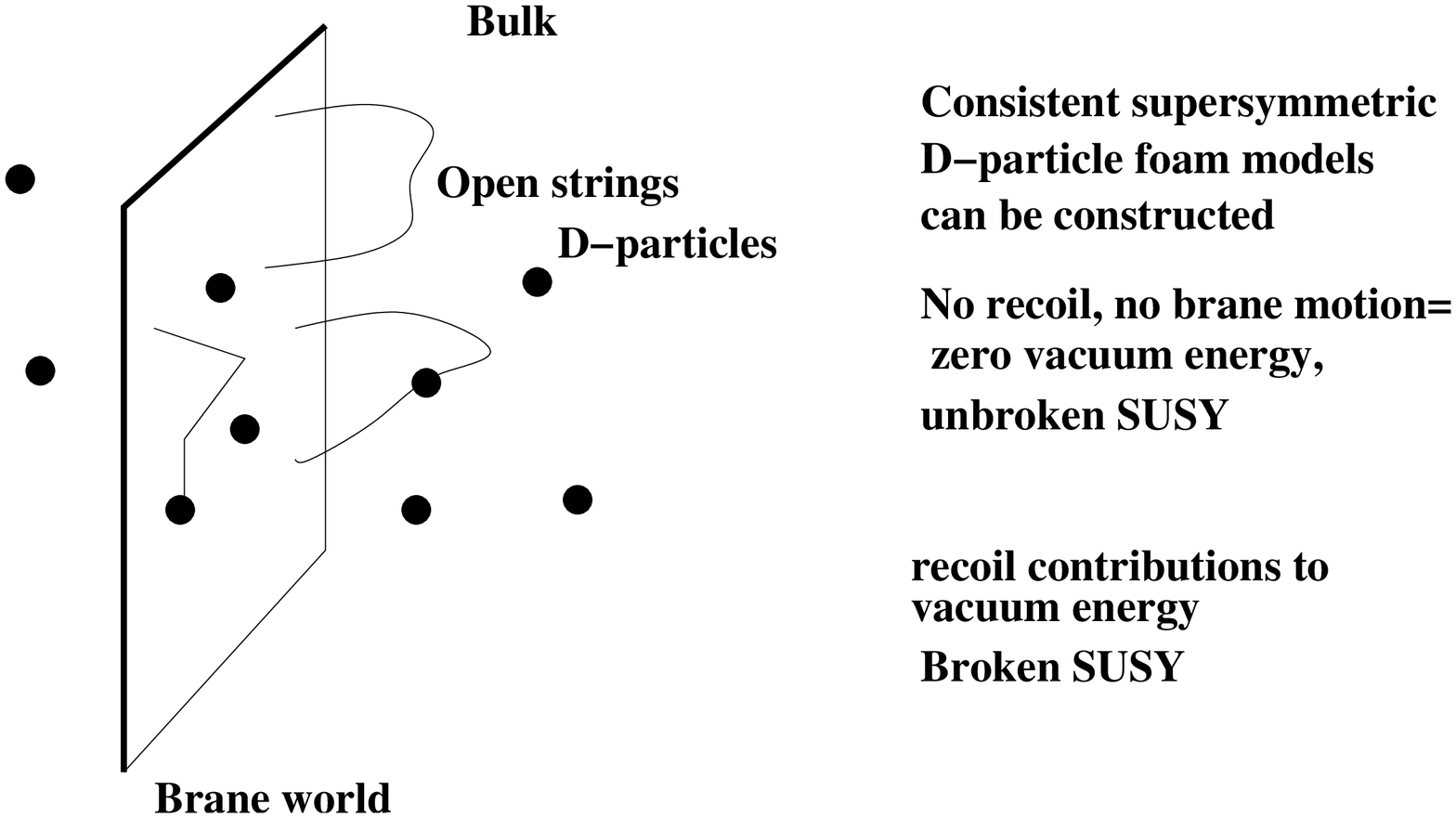} \caption{Schematic
representation of a D-foam. The figure indicates also the capture/recoil
process of a string state by a D-particle defect for closed (upper figure) and open
(lower figure) string states, in the presence of D-brane world. The presence of a
D-brane is essential due to gauge flux conservation, since an isolated
D-particle cannot exist. The intermediate composite state at $t=0$, which has
a life time within the stringy uncertainty time interval $\delta t$, of the
order of the string length, and is described by world-sheet logarithmic
conformal field theory, is responsible for the distortion of the surrounding
space time during the scattering, and subsequently leads to induced metrics
depending on both coordinates and momenta of the string state. This results on
modified dispersion relations for the open string propagation in such a
situation~\cite{Dfoam}, leading to \emph{non-trivial
optical properties} (refractive index\emph{ etc.}) for this space time.}%
\label{fig:recoil}%
\end{figure}

The model of space-time foam we are going to use in this work, is based on D-particles populating a bulk geometry between parallel
D-brane worlds. The model is termed D-foam~\cite{Dfoam} (c.f. figure
\ref{fig:recoil}), and our world is modelled as a three-brane moving in the
bulk geometry; as a result, D-particles cross the brane world and appear for
an observer on the brane as foamy structures which flash on and off .

Even at low energies $E$, such a foam may have observable consequences e.g.
decoherence  effects which may be of magnitude $O\left(  \left[  \frac
{E}{M_{P}}\right]  ^{n}\right)  $ with $n=1,2$, depending on the model, where $M_{P}$ is the Planck mass, or induced changes in the usual Lorentz invariant dispersion relations.
This results from topologically non-trivial interactions of the D-particles with the (open or closed) strings, involving splitting and capture of the latter by the D0-brane defects, as in fig.~\ref{fig:recoil}.

The study
of D-brane dynamics has been made possible by Polchinski's realisation~\cite{polch2} that
such solitonic string backgrounds can be described in a conformally invariant
way in terms of world sheets with boundaries \cite{polch2}. On these
boundaries Dirichlet boundary conditions for the collective target-space
coordinates of the soliton are imposed \cite{coll}. When low energy matter
given by a closed string propagating in a $\left(  d+1\right)  $-dimensional
space-time collides with a very massive D-particle (0-brane) embedded in this
space-time, the D-particle, due to its massive nature with mass $M_s/g_s$, where $M_s$ is the string scale, and $g_s < 1 $ is the (weak) string coupling, recoils~\cite{kogan} in a
non-relativistic manner. We shall consider the simple case of bosonic stringy
matter coupling to D-particles. Hence we can only discuss matters of principle
and ignore issues of stability due to tachyons. However we should note that an
open string model needs to incorporate for completeness, higher dimensional
D-branes such as the D3 brane. This is due to the vectorial charge carried by
the string owing to the Kalb-Ramond field. Higher dimensional D-branes (unlike
isolated D-particles~\cite{strominger}) can carry the charge from the endpoints of open strings that are
attached to them. It is for this reason, namely the conservation of the Kalb-Ramond flux, that in our D-particle foam we need the presence of higher-dimensional brane worlds embedded in the bulk (see figure \ref{fig:recoil}).

The current
state of phenomenolgical modelling of the interactions of D-particle foam with
stringy matter will be briefly summarised now. Since there are no rigid bodies
in general relativity the recoil fluctuations of the brane and the effective
stochastic back-reaction on space-time cannot be neglected.
To understand the formal structure of the world-sheet deformation operators
pertinent to the recoil/capture process, we first notice that
the world-sheet boundary operator $\mathcal{V}_{\rm{D}}$ describing the
excitations of a moving heavy D0-brane is given in the tree approximation by:
\begin{equation}
\mathcal{V}_{\rm{D}}=\int_{\partial D}\left(  y_{i}\partial_{n}X^{i}%
+u_{i}X^{0}\partial_{n}X^{i}\right)  \equiv\int_{\partial D}Y_{i}\left(
X^{0}\right)  \partial_{n}X^{i} \label{recoilop}%
\end{equation}
where $\partial D$ denotes the boundary of the world-sheet $D$ with the topology of a disk, to lowest order in string-loop perturbation theory,
$u_{i}$ and $y_{i}$ are the velocity \ and position of the D-particle
respectively and $Y_{i}\left(  X^{0}\right)  \equiv y_{i}+u_{i}X^{0}$. To
describe the capture/recoil we need an operator which has non-zero matrix elements between
different states of \ the D-particle and is turned on ``abruptly'' in target time. One way of doing this is to put~\cite{kogan} a $\Theta\left(  X^{0}\right)  $, the Heavyside function, in front of
$\mathcal{V}_{\rm{D}}$ which models an impulse whereby the D-particle starts
moving at $X^{0}=0$. This impulsive $\mathcal{V}%
_{\rm{D}}$, denoted by $\mathcal{V}_{\rm{D}}^{imp}$, can thus be represented
as
\begin{equation}
\mathcal{V}_{\rm{D}}^{imp}=\frac{1}{2\pi\alpha '}
\sum_{i=1}^{d}\int_{\partial D}d\tau\,u_{i}%
X^{0}\Theta\left(  X^{0}\right)  \partial_{n}X^{i}. \label{fullrec}%
\end{equation}
where $d$ in the sum denotes the appropriate  number of spatial target-space dimensions.
For a recoiling D-particle confined on a D3 brane, $d=3$.

Since $X^{0}$ is an operator it will be necessary to define $\Theta\left(
X^{0}\right)  $ as a \emph{regularized}  operator using the contour integral%
\begin{equation}
\Theta_{\varepsilon}\left(  X^{0}\right)  =-\frac{i}{2\pi}\int_{-\infty
}^{\infty}\frac{d\omega}{\omega-i\varepsilon}\e^{i\omega X^{0}}\rm{ with }\varepsilon
\rightarrow 0^+~,
\end{equation}
where $\varepsilon$ is a regulator, which, as discussed in \cite{kogan} and will be reviewed below, is linked with a running cutoff scale on the world-sheet of the string, on account of the requirement of the closure of the (logarithmic) conformal algebra.
Hence we can consider%
\begin{equation}\label{Depsilonop}
D_{\varepsilon}(X^0) \equiv D (X^0 ; \varepsilon) = X^{0}\Theta_{\varepsilon}\left(  X^{0}\right)
=-\int_{-\infty}^{\infty}\frac{d\omega}{\left(  \omega -i\varepsilon\right)  ^{2}%
}\e^{i\omega X^{0}}~.
\end{equation}
The introduction of the feature of impulse in the operator breaks conventional
conformal symmetry, but a modified logarithmic conformal algebra~\cite{lcft} holds~\cite{kogan}. A
generic logarithmic algebra in terms of operators $\mathcal{C}$ and
$\mathcal{D}$ and the stress tensor $T\left(  z\right)  $\ (in complex tensor
notation ) satisfies the operator product expansion
\begin{eqnarray}
T\left(  z\right)  \mathcal{C}\left(  w,\overline{w}\right)   &  \sim
\frac{\Delta}{\left(  z-w\right)  ^{2}}\mathcal{C}\left(  w,\overline
{w}\right)  +\frac{\partial\mathcal{C}\left(  w,\overline{w}\right)  }{\left(
z-w\right)  }+\cdots \nonumber \\
T\left(  z\right)  \mathcal{D}\left(  w,\overline{w}\right)   &  \sim
\frac{\Delta}{\left(  z-w\right)  ^{2}}\mathcal{D}\left(  w,\overline
{w}\right)  +\frac{1}{\left(  z-w\right)  ^{2}}\mathcal{C}\left(  w\right)
+\frac{\partial\mathcal{D}\left(  w\right)  }{\left(  z-w\right)  }+\cdots
\end{eqnarray}
and
\begin{eqnarray}
\left\langle \mathcal{C}\left(  z,\overline{z}\right)  \mathcal{C}\left(
0,0\right)  \right\rangle  &  \sim0\label{can1} \nonumber \\
\left\langle \mathcal{C}\left(  z,\overline{z}\right)  \mathcal{D}\left(
0,0\right)  \right\rangle  &  \sim\frac{c}{\left\vert z\right\vert ^{2\Delta}%
}\label{can2} \\
\left\langle \mathcal{D}\left(  z,\overline{z}\right)  \mathcal{D}\left(
0,0\right)  \right\rangle  &  \sim\frac{c}{\left\vert z\right\vert ^{2\Delta}%
}\left(  \log\left\vert z\right\vert +{c}\right)  \label{can3}%
\end{eqnarray}
where ${c}$ is a constant. Since the conformal dimension of
$\e^{iqX^{0}}$ is $\frac{q^{2}}{2}$ we find that
\begin{equation}
T\left(  w\right)  D_{\varepsilon}\left(  z\right)  \sim-\frac{\varepsilon
^{2}}{2\left(  w-z\right)  ^{2}}D_{\varepsilon}\left(  z\right)  +\frac
{1}{\left(  w-z\right)  ^{2}}\varepsilon\Theta_{\varepsilon}\left(
X^{0}\right)  +\cdots
\end{equation}
and so a logarithmic conformal algebra structure arises if we define
\begin{equation}\label{Cepsilonop}
C_{\varepsilon} (X^0) \equiv C(X^0 ; \varepsilon)  =\varepsilon\Theta_{\varepsilon}\left(
X^{0}\right)~,
\end{equation}
suppressing, for simplicity, the non-holomorphic piece. The
above logarithmic conformal field theory structure is found with this
identification. Similarly we find%
\[
T\left(  w\right)  C_{\varepsilon}\left(  z\right)  \sim-\frac{\varepsilon
^{2}}{2\left(  w-z\right)  ^{2}}C_{\varepsilon}\left(  z\right)  +\cdots
\]
Consequently $\Delta$ for $C_{\varepsilon}\left(  z\right)  $ and
$D_{\varepsilon}\left(  z\right)  $ is $-\frac{\varepsilon^{2}}{2}$. A
calculation (in a euclidean metric) for a disc of size $L$ with a
short-distance worldsheet cut-off $a$ reveals that as $\varepsilon\rightarrow0$%
\begin{eqnarray}
\left\langle C_{\varepsilon}\left(  z\right)  C_{\varepsilon}\left(  0\right)
\right\rangle  &  \sim&  O\left(  \varepsilon^{2}\right) \label{canep1}\\
\left\langle C_{\varepsilon}\left(  z,\overline{z}\right)  D_{\varepsilon
}\left(  0\right)  \right\rangle  &  \sim & \frac{\pi}{2}\sqrt{\frac{\pi
}{\varepsilon^{2}\alpha}}\left(  1-2\varepsilon^{2}\log\left\vert \frac{z}%
{a}\right\vert ^{2}\right) \label{canep2}\\
\left\langle D_{\varepsilon}\left(  z,\overline{z}\right)  D_{\varepsilon
}\left(  0\right)  \right\rangle  &  \sim & \frac{\pi}{2}\sqrt{\frac{\pi
}{\varepsilon^{2}\alpha}}\left(  \frac{1}{\varepsilon^{2}}-2\log\left\vert
\frac{z}{a}\right\vert ^{2}\right)  \label{canep3}%
\end{eqnarray}
where $\alpha=\log\left\vert \frac{L}{a}\right\vert ^{2}$. We consider
$\varepsilon\rightarrow0+$ such that
\begin{equation}
\varepsilon^{2}\alpha\sim\frac{1}{2\eta}=O\left(  1\right)  ~,
\label{epscutoff}%
\end{equation}
where $\eta$ is the time signature and the right-hand side is kept fixed as
the cutoff runs; it is then straightforward to see that (\ref{canep1}),
(\ref{canep2}), and (\ref{canep3}) are consistent with (\ref{can1}),
(\ref{can2}), and (\ref{can3}). It is only under the condition
(\ref{epscutoff}) that the recoil operators $C_{\varepsilon}$ and
$D_{\varepsilon}$ obey a closed logarithmic conformal algebra~\cite{kogan}:
\begin{eqnarray}
<C_{\varepsilon}(z)C_{\varepsilon}(0)>  &  \sim & 0\nonumber\\
<C_{\varepsilon}(z)D_{\varepsilon}(0)>  &  \sim & 1\nonumber\\
<D_{\varepsilon}(z)D_{\varepsilon}(0)>  &  \sim & -2\eta\log|z/L|^{2} \label{CD}%
\end{eqnarray}
The reader should notice that the full recoil operators, involving
$\partial_{n}X^{i}$ holomorphic pieces with the conformal-dimension-one
entering (\ref{fullrec})), obey the full logarithmic algebra (\ref{can1}),
(\ref{can2}), (\ref{can3}) with conformal dimensions $\Delta=1-\frac
{\varepsilon^{2}}{2}$. From now on we shall adopt the Euclidean signature
$\eta=1$.

We next remark that, at tree level in the string perturbation sense, the
stringy sigma model (inclusive of the D-particle \ boundary term and other
\ vertex operators) is a two dimensional renormalizable quantum field theory;
hence for generic couplings $g^{i}$ it is possible to see how the couplings
run in the renormalization group sense with changes in the short distance
cut-off through the beta functions $\beta^{i}$. In the world-sheet
renormalization group \cite{klebanov}, based on expansions in powers of the
couplings, $\beta^{i}$ has the form ( with no summation over the repeated
indices)
\begin{equation}
\beta^{i}=y_{i}g^{i}+\ldots\label{beta fn}%
\end{equation}
where $y_{i}$ is the anomalous dimension, which is related to the conformal
dimension $\Delta_{i}$ by $y_{i} = \Delta_{i} - \delta$, with $\delta$ the
engineering dimension (for the holomorphic parts of vertex operators for the
open string $\delta= 1$). The $\ldots$ in (\ref{beta fn}) denote higher orders
in $g^{i}$. Consequently, in our case, we note that the (renormalised)
D-particle recoil velocities $u^{i}$ constitute such $\sigma$-model couplings,
and to lowest order in the renormalised coupling $u_{i}$ the corresponding
$\beta$ function satisfies%
\begin{equation}
\frac{du^{i}}{d\log\Lambda}=-\frac{\varepsilon^{2}}{2}u^{i}. \label{rengp}%
\end{equation}
where $\Lambda$ is a (covariant) world-sheet renormalization-group scale.
In our notation, we
identify the logarithm of this scale with $\alpha=\log\left\vert \frac{L}%
{a}\right\vert ^{2}$, satisfying (\ref{epscutoff}).

An important comment is now in order concerning the interpretation of the flow
of this world-sheet renormalization group scale as a target-time flow. The
target time $t$ is identified through $t=2\log\Lambda$. For completeness we
recapitulate the arguments of \cite{kogan} leading to such a conclusion. Let
one make a scale transformation on the size of the world-sheet
\begin{equation}
L\rightarrow L^{\prime}=\e^{t/4}L \label{fsscaling}%
\end{equation}
which is a finite-size scaling (the only one which has physical sense for the
open string world-sheet). Because of the relation between $\varepsilon$ and
$L$ (\ref{epscutoff}) this transformation will induce a change in
$\varepsilon$
\begin{equation}
\varepsilon^{2}\rightarrow\varepsilon^{\prime2}=\frac{\varepsilon^{2}%
}{1+\varepsilon^{2}t} \label{epsilontransform}%
\end{equation}
(note that if $\varepsilon$ is infinitesimally small, so is $\varepsilon^{\prime}$
for any finite $t$). From the scale dependence of the correlation functions
(\ref{CD}) that $C_{\varepsilon}$ and $D_{\varepsilon}$ transform as:
\begin{eqnarray}\label{cdtrans}
D_{\varepsilon}  &  \rightarrow  & D_{\varepsilon^{\prime}}=D_{\varepsilon
}+tC_{\varepsilon}\nonumber\\
C_{\varepsilon}  &  \rightarrow  & C_{\varepsilon^{\prime}}=C_{\varepsilon}%
\end{eqnarray}
From this transformation one can then see that the coupling constants in front
of $C_{\varepsilon}$ and $D_{\varepsilon}$ in the recoil operator
(\ref{recoilop}), i.e. the velocities $u_{i}$ and spatial collective
coordinates $y_{i}$ of the brane, must transform like:
\begin{equation}
u_{i}\rightarrow u_{i}~~,~~y_{i}\rightarrow y_{i}+u_{i}t \label{scale2}%
\end{equation}
This transformation is nothing other but the Galilean
transformation for the heavy D-particles and thus it demonstrates that the finite size scaling parameter $t$, entering
(\ref{fsscaling}), plays the r\^ole of target time, on account of (\ref{epscutoff}). Notice that (\ref{scale2}) is derived upon using (\ref{CD}), that is in the limit where $\varepsilon \to 0$.
This will become important later on, where we shall discuss (stochastic) relaxation phenomena in our recoiling D-particle.

Thus, in the presence of recoil a world-sheet scale transformation leads to an
evolution of the $D$-brane in target space, and from now on we identify the
world-sheet renormalization group scale with the target time $t$. In this
sense, Eq.~(\ref{rengp}) is an evolution equation in target time.

\subsection{World-sheet genus summation and quantum fluctuations of recoil velocity\label{subsec:genera}}

However, Eq.~(\ref{rengp}) does not capture quantum-fluctuation aspects of $u^{i}$ about its classical trajectory with time $u_i(t)$. Going
to higher orders in perturbation theory of the quantum field theory at fixed
genus does not qualitatively alter the situation in the sense that the
equation remains deterministic. In the next section we shall consider the
effect of string perturbation theory where higher genus surfaces are
considered and re-summed in some appropriate limits that we shall discuss in detail.

It is not possible to exactly sum up higher orders in string perturbation
theory. We have seen that infrared singularities in the integration over the
moduli of the Riemann surface (representing the world sheet) in the wormhole
limit are related to the recoil operators for the D-particle. The wormhole
construction \cite{coleman} is a way of constructing higher genus surfaces
from lower genus ones. Since it will be relevant to us later, we should note
that $g_{s}$ the string coupling is given by
\begin{equation}
g_{s}=\mathrm{e}^{\left\langle \Phi\right\rangle } \label{string coupling}%
\end{equation}
where $\Phi$ is the spin zero dilaton mode which is part of \ the massless
string multiplet. Here $\left\langle \ldots\right\rangle $ denotes the string
path integral $\int DX\,\mathrm{e}^{S_{\sigma}}\Phi$ where $S_{\sigma}$ is the
string $\sigma$-model action in the presence of string backgrounds such as the
dilaton and the Kalb-Ramond modes. In particular the $\sigma$ model
deformation due to the dilaton has the form
\begin{equation}
\frac{1}{4\pi}\int_{\Sigma}d\sigma d\tau\,\sqrt{\gamma}\,\Phi\left(  X\right)
R^{\left(  2\right)  }\left(  \tau,\sigma\right)  \label{dilaton}%
\end{equation}
on a worldsheet Riemann surface $\Sigma$ where $\gamma_{\alpha\beta}$ is the
induced metric on the worldsheet, $\gamma=\left\vert \det\gamma_{\alpha\beta
}\right\vert $ and \ $R^{\left(  2\right)  }$ is the associated Ricci
curvature scalar. Now the Euler characterisitic $\chi$ of $\Sigma$ is given by%
\begin{equation}
\chi=\frac{1}{4\pi}\int_{\Sigma}d\sigma d\tau\sqrt{\gamma}R^{\left(  2\right)
}=2\left(  1-g\right)  \label{euler}%
\end{equation}
where $g$ is the genus and is an integer valued invariant. If we split the
dilaton into a classical (worldsheet co-ordinate independent) part
$\left\langle \Phi\right\rangle $ and a quantum part $\varphi=\colon\Phi
\colon$, where $\colon\ldots\colon$ denotes appropriate normal ordering, we
can write $\Phi=\left\langle \Phi\right\rangle +\varphi$. The $\sigma$-model
partition function $Z$\ can be written as a sum over genera
\begin{eqnarray}
Z  &  = & \sum_{\chi}\int\int d\gamma_{\alpha\beta}\,dX\,\mathrm{e}%
^{-S_{{\rm rest}}-\chi\left\langle \Phi\right\rangle -\frac{1}{4\pi}\int_{\Sigma
}d\sigma d\tau\,\sqrt{\gamma}\,\varphi R^{\left(  2\right)  }\left(
\tau,\sigma\right)  }\label{partiition1}\\
&  = & \sum_{\chi}g_{s}^{-\chi}\int\int d\gamma_{\alpha\beta}\,dX\,\mathrm{e}%
^{-S_{rest}-\frac{1}{4\pi}\int_{\Sigma}d\sigma d\tau\,\sqrt{\gamma}\,\varphi
R^{\left(  2\right)  }\left(  \tau,\sigma\right)  } \label{partition2}%
\end{eqnarray}
where $S_{{\rm rest}}$ denotes a $\sigma$-model action involving the rest of the
background deformations except the dilaton. For the moment we will assume that
the theory is such that a potential is generated for $\Phi$ which suppresses
the fluctuations represented by $\varphi$. In general we would have to
consider $g_{s}=\e ^{\Phi}$ which would then make the string coupling a field.

The summation over genera cannot be performed exactly. We will follow an
approach using a mechanism due to Fischler and Susskind \cite{fs},\cite{szabo}
based on a dilute gas of wormholes (proposed originally by Coleman within the
context of Euclidean quantum gravity \cite{coleman}). This results in the
structure of recoil (in lowest order) being modified by generating a gaussian
distribution for the recoil velocity $u^{i}$.

A detailed review of the pertinent formalism has been given in \cite{szabo} and will not be repeated here. For our
purposes in this section we only note that, in the case of mixed logarithmic
states, the pinched topologies are characterized by divergences of a double
logarithmic type which arise from the form
of the string propagator in the presence
of generic logarithmic operators $C$ and $D$,
$\int dq~q^{\Delta_{\varepsilon
}-1}\,\langle C,D|%
\left(\begin{array}{cc} 1 \quad \log q \\ 0  \quad 1
\end{array}\right) |C,D\rangle$~.

As shown in \cite{kogan}, the mixing between $C$ and $D$ states along
degenerate handles on the world-sheet (c.f. figure \ref{fig:dbranes}) leads formally to divergent string propagators in physical
amplitudes, whose integrations have leading divergences of the form
\begin{equation} \int
\frac{dq}q~\log q\int d^{2}z~D(z;\varepsilon)\int d^{2}z^{\prime}~C(z^{\prime2
}\int d^{2}z~D(z;\varepsilon)\int d^{2}z^{\prime}~C(z^{\prime};\varepsilon
)~.\label{bilocal}
\end{equation}
As explained in \cite{szabo}, these $(\log\delta)^{2}$ divergences can be cancelled by imposing momentum conservation in the scattering process of the light string states off
the D-particle background.

We stress again at this stage that isolated D-particles do not
exist, as a result of their gauge flux conservation requirement. The
physically correct way to formulate, therefore, the problem, is to consider
groups of $N$, say, D-particles, which interact among themselves with
flux-carrying stretched strings. The above analysis remains intact (in the
sense of generalising straightforwardly) when more than one D-particle is
present in a region with typical dimensions smaller than the string length
sometimes known as the fat brane \cite{szabo}. There are, of course, technical
differences in the sense that a non-abelian structure arises and the
$\widehat{Y}$ have matrix labels. All qualitative features, however, are
preserved concerning the quantisation of the D-particle background moduli. For
the remainder of this section we shall, therefore, formulate our arguments
within this rigorous multi-D-particle picture.

The cancelation of leading divergences of the genus expansion in the
non-abelian case of a group of $N$ D-particles, has been demonstrated explicitly in
\cite{szabo}. It is shown there that this renormalization requires that the
change in (renormalized) velocity of the D-particle, due to the recoil from the scattering of string states, be
\begin{eqnarray}
\bar u_{i}^{ab}=-\frac1{M_{D}}\,\Bigl(k_{1}+k_{2}\Bigr)_{i}\,\delta^{ab}%
=\frac{d\bar Y_{i}^{ab}}{dt}~, \qquad a,b = 1 \dots N \label{recoilvel}%
\end{eqnarray}
where $k_{1,2}$ are the initial and final momenta in the scattering process
and $M_{D}=1/\sqrt{\alpha^{\prime}}\, g_{s}$ is the BPS mass of the string
soliton \cite{polch,polch2}, and $g_{s} < 1 $ is the physical (weak) string
coupling. In (\ref{recoilvel}), the $k_{1,2}$ are true physical momenta so
that $M_{D}$ represents the actual BPS mass of the D-particles. This means
that, to leading order, the constituent D-particles in a group of $N$ of them,
say, move parallel to one another with a common velocity and there are no
interactions among them. Thus the leading recoil effects imply a commutative
structure and the ``fat brane'' of the group of D-particles behaves as a
single D-particle (with a single average collective coordinate of its center
of mass). In such a limit one may replace $u_{i}^{ab}$ by $u_{i}$ (c.f.
section \ref{subsec:lcft}), describing the collective recoil velocity of the fat brane.
This should be understood throughout this work.

In addition to this divergence, there are sub-leading $\log\delta$
singularities, corresponding to the diagonal terms
$$\int d^{2} z~D(z;\varepsilon)\int d^{2}z^{\prime}~D(z^{\prime};\varepsilon) \quad {\rm and} \quad \int
d^{2}z~C(z;\varepsilon)\int d^{2}z^{\prime}~C(z^{\prime};\varepsilon)~.$$ These
latter terms are the ones we should concentrate upon for the purposes of
deriving the quantum fluctuations of the collective D-particle coordinates. It
is these sub-leading divergences in the genus expansion which lead to
interactions between the constituent D-branes and provide the appropriate
\emph{noncommutative quantum extension} of the leading dynamics (\ref{recoilvel}).
The reader should recall that these (sub-leading) divergences also showed up
in the much simpler case of perpetual Galilean motion of D-branes discussed in
\cite{coll}, as a result of the translational symmetries
zero mode contributions.

In the weak-coupling case, we can truncate the genus expansion to a sum over
pinched annuli (fig. \ref{fig:dbranes}). This truncation
corresponds to a semi-classical approximation to the full quantum string
theory in which we treat the D-particles as heavy non-relativistic objects in
target space. Then the dominant contributions to the sum are given by the
$\log\delta$ modular divergences described above, and the effects of the
dilute gas of wormholes on the disc are to \emph{exponentiate} the bilocal
operator (\ref{bilocal}), describing string propagation in a
pinched annulus. Thus, in the pinched approximation, the genus expansion of
the bosonic $\sigma$-model leads to an effective change in the matrix $\sigma
$-model action by~\cite{szabo}
{\small \begin{eqnarray}
\Delta S\simeq\frac{g_{s}^{2}}2\log\delta\sum_{a,b,c,d}\,\int_{-\infty
}^{\infty}d\omega~d\omega^{\prime}~\oint_{\partial\Sigma}\oint_{\partial
\Sigma^{\prime}} V_{ab}^{i}(x;\omega)~G_{ij}^{ab;cd}(\omega,\omega^{\prime
})~V_{cd}^{j}(x;\omega^{\prime}) \label{actionchange}%
\end{eqnarray}}
where $\omega, \omega^{\prime}$ are Fourier variables, defined appropriately
in \cite{szabo}, and $G_{ij}$~, $i,j = C,D$ is a metric in the theory space of
strings, introduced by Zamolodchikov~\cite{zam}.

The bilocal action (\ref{actionchange}) can be cast into the form of a local
worldsheet effective action by using standard tricks of wormhole calculus
\cite{coleman} and rewriting it as a functional Gaussian
integral~\cite{szabo}
{\small \begin{eqnarray}
&& \mathrm{e}^{\Delta S}   =  \int[d\breve{\rho}]~\exp\left[  -\frac
12\sum_{a,b,c,d}\,\int_{-\infty}^{\infty}d\omega~d\omega^{\prime}~\breve{\rho
}_{i}^{ab}(\omega)~\oint_{\partial\Sigma} \oint_{\partial\Sigma^{\prime}%
}G^{ij}_{ab;cd}(\omega,\omega^{\prime})~ \breve{\rho}_{j}^{cd}(\omega^{\prime
})\right. \nonumber\\
&&  \left.  ~~~~~~~~~~~~~~~~~~~~ +\,g_{s}\,\sqrt{\log\delta}~\sum_{a,b=1}%
^{N}\,\int_{-\infty}^{\infty}d\omega~\breve{\rho}_{i}^{ab}(\omega
)\,\oint_{\partial\Sigma} V_{ab}^{i}(x;\omega)\right]  \label{Gaussianint}%
\end{eqnarray}}where $\breve{\rho}_{i}^{ab}(\omega)$ are stochastic coupling constants of the
worldsheet matrix $\sigma$-model, which express quantum fluctuations of the
corresponding background fields in target space, as a consequence of genus
re-summation. Thus the effect of the resummation over pinched genera is to
induce quantum fluctuations of the collective D-brane background, leading to a
set of effective quantum coordinates
{\begin{eqnarray}
\breve{Y}_{i}^{ab}(\omega)~\to~\widehat{\mathcal{Y}}_{i}^{ab}(\omega
)=\breve{Y}_{i}^{ab}(\omega)+g_{s}\,\sqrt{\log\delta}~ \breve{\rho}_{i}%
^{ab}(\omega) \label{qcoupling}%
\end{eqnarray}}
viewed as position operators in a co-moving target space frame.

Thus we find that the genus expansion in the pinched approximation for the bosonic string is~\cite{szabo}
{\begin{eqnarray}
\sum_{h^{(p)}}Z_{N}^{h^{(p)}}[A]~\simeq~\left\langle \int_{\mathcal{M}}%
[d\rho]~\wp[\rho] {}~W\!\left[  \partial\Sigma
;A-\mbox{$\frac1{2\pi\alpha'}$}\,\rho\right]  \right\rangle _{0}
\label{pinchedpartfn}%
\end{eqnarray}}
where the sum is over all pinched genera of infinitesimal pinching size, and
{\small \begin{eqnarray}
\wp[\rho] \propto\exp\left[  -\frac1{2\Gamma^{2}}\sum_{a,b,c,d}\,\int_{0}%
^{1}ds~ds^{\prime}~ \rho_{i}^{ab}\left(  X^{0}(s)\right)  \,G_{ab;cd}%
^{ij}(s,s^{\prime})\,\rho_{j}^{cd} \left(  X^{0}(s^{\prime})\right)  \right]
\label{Gaussiandistr}%
\end{eqnarray}}
is a (appropriately normalized) functional Gaussian distribution on moduli
space of width
\begin{eqnarray}
\Gamma=g_{s}\,\sqrt{\log\delta} \label{widthdef}%
\end{eqnarray}
In (\ref{pinchedpartfn}) we have normalized the functional Haar integration
measure $[d\rho]$ appropriately.

We see therefore that the diagonal sub-leading logarithmic divergences in the
modular cutoff scale $\delta$, associated with degenerate strips in the genus
expansion of the matrix $\sigma$-model, can be treated by absorbing these
scaling violations into the width $\Gamma$ of the probablity distribution
characterizing the quantum fluctuations of the (classical) D-brane
configurations $Y_{i}^{ab}(X^{0}(s))$. In this way the interpolation among
families of D-brane field theories corresponds to a quantization of the
worldsheet renormalization group flows. Note that the worldsheet wormhole
parameters, being functions on the moduli space of recoil deformations, can be
decomposed as
\begin{eqnarray}
\rho_{i}^{ab}(X^{0}(s))=\lim_{\varepsilon\to0^{+}}\left(  [\rho_{C}]_{i}%
^{ab}C(X^{0};\varepsilon) +[\rho_{D}]_{i}^{ab}D(X^{0};\varepsilon)\right)
\label{wormholedecomp}%
\end{eqnarray}
The fields $\rho_{C,D}$ are then renormalized in the same way as the D-brane
couplings, so that the corresponding renormalized wormhole parameters generate
the same type of (Galilean) $\beta$-function equations (\ref{rengp}).

According to the standard Fischler-Susskind mechanism for canceling string
loop divergences~\cite{fs}, modular infinities should be identified with
worldsheet divergences at lower genera. Thus the strip divergence $\log\delta$
should be associated with a worldsheet ultraviolet cutoff scale $\log\lambda$,
which in turn is identified with the target time as described earlier.

We may in effect take $\delta$ independent from $\Lambda$, in which case we
can first let $\varepsilon\to0^{+}$ in the above and then take the limit
$\delta\to0$. Interpreting $\log\delta$ in this way as a renormalization group
time parameter (interpolating among D-brane field theories), the time
dependence of the renormalized width (\ref{widthdef}) expresses the usual
properties of the distribution function describing the time evolution of a
wavepacket in moduli space. The inducing of a statistical Gaussian spread of
the D-brane couplings is the essence of the quantization procedure.

A final remark is in order. From the form (\ref{Depsilonop}) and
(\ref{Cepsilonop}) of the recoil operators, it is evident that the dominant
contributions in the limit $\varepsilon\to0^{+}$, we consider here, come from
the $D$-deformations, pertaining to the recoil velocity $u^{i}$ of the
D-particle (or, better, the center of mass velocity of a group of D-particles,
as discussed above). From now on, therefore, we restrict our attention to the
distribution functions of such recoil velocities:
\begin{equation}
\wp(u) \sim\frac{1}{\Gamma} \mathrm{e}^{-\frac{u^{2} - {\bar u}^{2}}%
{\Gamma^{2}}}~, \qquad\Gamma= g_{s} \sqrt{\mathrm{log}\delta}~,
\label{gaussian}%
\end{equation}
where $\bar u$ denotes the classical recoil velocity. Notice that, upon
invoking~\cite{szabo} the Fischler-Susskind mechanism~\cite{fs} for the
absorption of the modular infinities to lower-genus (disc) world-sheet
surfaces, we may identify $\mathrm{log}\delta$ with the target time:
\begin{equation}
\mathrm{log}\delta= t~, \label{modtime}%
\end{equation}
where this identification should be understood as being implemented at the end
of the computation. To be precise, as explained in \cite{szabo}, the correct
form of (\ref{modtime}) would be: $\mathrm{log}\delta= g_{s}^{\chi}~ t$, with
$\chi> 0$ an exponent that can only be determined phenomenologically in the
approach of \cite{szabo}, by comparing the space-time uncertainty principles,
derived in this approach of re-summing world-sheet genera, with the ones
within standard string/brane theory. In fact, in our approach of re-summing
world-sheet pinched surfaces~\cite{szabo}, one obtains for the spatial and
temporal variances:
\begin{equation}\label{yoneunc}
\Delta Y^{aa} \Delta t \ge g_{s}^{\chi}\sqrt
{\alpha^{\prime}}~,
\end{equation}which implies that the standard string-theory
result for the space-time uncertainty relation~\cite{yoneya}, independent of the string coupling, is obtained for $\chi= 0$. This is the case we shall consider here, which leads to the
identification (\ref{modtime}). However, in the modern approach of D-brane
theories, one can adjust the uncertainty relations in order to probe minimal
distances below the string length, which is achieved by the
choice~\cite{szabo}, e.g. $\chi= 2/3$, reproducing the characteristic minimal
length probed by D-particles~\cite{liyoneya}. In our case, where, as we shall
discuss in the next subsection, the coupling constant of the string may itself
fluctuate, it is the mean value of $g_{s}$ that enters in such relations. This
issue is not relevant if we stay within the $\chi= 0$ case, which we do in
this article.

We next remark that the nature of the Gaussian correlation is assumed to be
delta correlated in time. The Langevin equation \cite{gardiner2} implied by
(\ref{qcoupling}) replaces (\ref{rengp}) and can be written as
\begin{equation}\label{langevin}
\frac{d{\bar{u}}^{i}}{dt}=-\frac{1}{4t}{\bar{u}}^{i}\,+\frac{g_{s}}%
{\sqrt{2\alpha^{\prime}}}t^{1/2}\xi\left(  t\right)  \label{Langevin}%
\end{equation}
where $t=\varepsilon^{-2}$ and $\xi\left(  t\right)  $ represents white noise.
This equation is valid for large $t$. From the above analysis it is known that
\cite{szabo} to $O\left(  g_{s}^{2}\right)  $ the
correlation for $\xi\left(  t\right)  $ is $\bar{u}^{i}$ independent and,
for time scales of interest, is correlated like white noise ; hence the
correlation of $\xi\left(  t\right)  $ has the form:
\begin{equation}
\left\langle \xi\left(  t\right)  \xi\left(  t^{\prime}\right)  \right\rangle
=\delta\left(  t-t^{\prime}\right)  .
\end{equation}
Since the vectorial nature of $\bar{u}^{i}$ is not crucial for our analysis we
will suppress it and consider the single variable $\bar{u}$.
Eq.~(\ref{langevin}) is a stochastically fluctuating quantum equation that replaces the ordinary renormalization-group equation (\ref{rengp}) (with (\ref{epscutoff}) being assumed), valid at tree-level on the world-sheet genus expansion.

We should stress that this equation is valid for large $t$ which is required
since $\varepsilon$ is small. Hence the apparent singularity in
Eq.~(\ref{Langevin} ) at $t=0$ is not relevant and so we can empirically
\emph{regularise} this singularity by changing $\frac{1}{t}$ to $\frac
{1}{t+t_{0}}$ for some $t_{0}>0$; $t_{0}$ is the order of the capture time of
the matter string by the D-particle. The stochastic Langevin equation
(\ref{Langevin}), describes relaxation aspects of the recoiling D-particle
with equilibrium being reached only as $\varepsilon\rightarrow 0$ (or
$t\rightarrow\infty$).

The reader should notice that in the limit the system reaches equilibrium with
a constant in time velocity. It is only in this limit that the Galilean
transformation (\ref{scale2}) applies, as already discussed there. We now
proceed to a solution of this Langevin equation and a discussion on the
pertinent physical consequences for a statistical population of
quantum-fluctuating D-particles~\cite{tsallissarkar}.

\subsection{Solution of (quantum) Langevin equation for recoil velocity \label{subsec:langevin}}

Eq.~(\ref{Langevin}) is \ particulary
simple equation in the sense that the drift and diffusion terms are
independent of $u$. By making a change of variable it is easy to eliminate the
drift term and the resulting equation can then be interpreted in terms of a
Wiener process~\cite{gardiner2}. Let us consider the auxiliary equation
\begin{equation}
\frac{d}{dt}y=-\frac{1}{4\left(  t+t_{0}\right)  }y, \label{drift}%
\end{equation}
which just deals with the drift part of Eqn.(\ref{Langevin}). It has a
solution
\[
y\left(  t\right)  =y\left(  t_{0}\right)  \Upsilon\left(  t\right)
\]
where
\begin{equation}
\Upsilon\left(  t\right)  =\exp\left[  -\frac{1}{4}\int_{0}^{t}\frac
{dt^{\prime}}{t^{\prime}+t_{0}}\right]  =\left(  \frac{t+t_{0}}{t_{0}}\right)
^{-\frac{1}{4}}%
\end{equation}
and $t_{0}$ is a time much smaller than $t$. We now define $U\left(  t\right)
=u\left(  t\right)  \Upsilon\left(  t\right)  ^{-1}$ and readily find that
\begin{equation}
\frac{dU}{dt}=\frac{g_{s}}{\sqrt{2\alpha^{\prime}}}t^{1/2}\Upsilon\left(
t\right)  ^{-1}\xi\left(  t\right)  . \label{Langevin2}%
\end{equation}
This describes purely diffusive motion and is thus related to the Wiener
process; equivalently we can consider the associated probability distribution
$p\left(  U,t\right)  $ which satisfies the Fokker-Planck equation%
\begin{equation}
\frac{\partial}{\partial t}p\left(  U,t\right)  =\frac{1}{4\alpha^{\prime}%
}g_{s}^{2}t(\frac{t+t_{0}}{t_{0}})^{\frac{1}{2}}\frac{\partial^{2}}{\partial
U^{2}}p\left(  U,t\right)  . \label{Fokker-Planck}%
\end{equation}
If at $t=0$ consider a D-particle velocity recoil $u_{0}$ so that
\begin{equation}
p\left(  U,0\right)  =\delta\left(  U-u_{0}\right)  ~. \label{initial}%
\end{equation}
Eq.~(\ref{Fokker-Planck}) can be solved to give
\begin{equation}
p\left(  U,t\right)  =\sqrt{\frac{15\alpha^{\prime}}{2\pi\eta\left(  t\right)
}}\frac{1}{g_{s}}\exp\left(  -\frac{15\alpha^{\prime}(U-u_{0})^{2}}{2g_{s}%
^{2}\eta\left(  t\right)  }\right)  \label{gaussian1}%
\end{equation}
where%
\begin{equation}
\eta\left(  t\right)  =2t_{0}^{2}+3\left(  t+t_{0}\right)  ^{2}\sqrt{1+\frac
{t}{t_{0}}}-5t_{0}^{\frac{1}{2}}(t+t_{0})^{\frac{3}{2}}. \label{dispersion}%
\end{equation}
If the D-particle is typically interacting with matter on time scales of
$t_{0}$, then the effect of a large number of such collisions can be
calculated by performing an ensemble average over a distribution of $u_{0}$. A
distribution for $u_{0}$ that has been used in modelling is a gaussian with
zero mean and variance $\sigma$. This is readily seen to lead to an averaged
distribution D-particle velocity recoil distribution $\langle p\left(  u\left\vert
g_{s}\right.  \right)  \rangle$ where
\begin{eqnarray}\label{urecflct}
 \langle p\left(  u\left\vert g_{s}\right.  \right)  \rangle
=\sqrt{\frac{15\alpha^{\prime}}{2\pi\left(  g_{s}^{2}\eta\left(  t\right)
+15\alpha^{\prime}\sigma^{2}\right)  }}\exp\left[  -\frac{15\alpha^{\prime}%
}{2\left(  g_{s}^{2}\eta\left(  t\right)  +15\alpha^{\prime}\sigma^{2}\right)
}u^{2}\right]  ~.
\end{eqnarray}
 We have used a notation for $p$ which emphasises that it is conditional on
$g_{s}$ having a fixed value. This is due to the possibility that in string theory
the string coupling itself may fluctuate, since it is given by the exponential of the v.e.v. of the dilaton field, $\Phi$, (\ref{string coupling}), which in turn may fluctuate according to some distribution (``fuzzy''), as it depends in general on target-spacetime  properties, such as temperature $T$ or, as is the case of the D-particle foam considered here, topologically non-trivial fluctuations of space-time \emph{etc}.
This will lead~\cite{tsallissarkar} to non-extensive statistics of the quantum fluctuating recoil velocities (that depend on $g_s$ (c.f. Eq.~(\ref{recoilvel})). However, because such additional fuzzyness is small compared to the standard effects of genus summation and statistical properties of the foam for the range of energies of physical interest~\cite{tsallissarkar}, we shall not discuss such super-statistics properties of the D-particle foam further in this work.

We next remark that the interaction time entering (\ref{urecflct})
includes \emph{both} the time for capture and re-emission of the string by the
D-particle, as well as the time interval until the next capture, during string propagation.
In a generic situation, this latter time interval could be much larger than
the capture time, especially in dilute gases of D-particles, which include less than one D-particle per string ($\alpha^{3/2}$) volume. Indeed, as discussed
in detail in \cite{emnuncertnew}, using generic properties of strings
consistent with the space-time uncertainties~\cite{yoneya}, the capture and
re-emission time $t_{0}$, involves the growth of a stretched
string between the string state and the D-brane world (c.f.
fig.~\ref{fig:recoil}) and is found proportional to the incident string energy
$E$:
\begin{equation}
t_{0} \sim\alpha^{\prime}\, E  \ll \sqrt{\alpha ^{\prime}}~.
\end{equation}
In view of (\ref{urecflct}), then, averaging over the capture time yields a
good estimate of the order of magnitude of the quantum fluctuations  $\langle u_i u_j \rangle \equiv \sigma_0 \delta_{ij} $ of the recoil velocity in the D-particle foam model:
\begin{equation}
\sigma_0 \sim \left(\frac{E}{M_s}g_s\right)^2~.
\label{recvelquantuflct}
\end{equation}
We shall return to this estimate when we discuss the order of magnitude of decoherence and CPT-violating effects in D-particle foam, in the next subsection.

An important remark is in order here, concerning the nature of matter probes to be used in testing such effects.
As a result of the capture process of fig.~\ref{fig:recoil}, only \emph{electrically neutral} matter probes~\cite{aes,emnuncertnew} can be captured by the electrically neutral D-particle defects, for reasons of electric charge conservation, which is assumed to be an exact symmetry in any quantum gravity or string theory model. Thus, for electrons, for instance, the D-particle foam will appear transaparent, but not for photons, neutrinos, and in general neutral particles (such as neutral mesons etc.). In this latter respect, we should mention the following: at a string theory level, the quark (electrically charge) constituents of the neutral meson will be transparent, according to the above argument, to the D-particle foam, but this will not be the case for gluons. In this sense the neutral meson will feel the effects of the D-particle foam, but the strong interaction effects might affect (suppress) the strength of the phenomenon. The neutrino, on the other hand, being an elementary string excitation, will not suffer from such suppressions.
This are important features to be taken seriously into account when one searches experimentally for such effects.

\subsection{Quantum Decoherence due to recoil-velocity fluctuations \label{subsec:decoh}}

The above-described induced stochastic fluctuations of the D-particle recoil velocity $u_i$, as a result of genus summation in the world-sheet of the string, imply \emph{canonical quantisation }for $u_i$, which in this sense is viewed in target space as a quantum operator~\cite{emn}. Indeed, as discussed in detail in \cite{emninfl,szabo}
the interpretation of the target time as a world-sheet renormalization group scale is compatible with the Helmholtz conditions required for canonical quantisation of a dynamical system. In other words,  the derivation of the dynamical target-space equations describing D-particle recoil can be shown rigorously to be derivable from an effective action, not necessarily on-shell though. This off-shellness is related to the deviation from conformal invariance, as a result of the (small but finite) anomalous dimension $-\varepsilon^2/2$, as discussed in sub-section \ref{subsec:lcft}, c.f. Eqs.~(\ref{rengp}), (\ref{epscutoff}).

The non-conformal nature of the recoiling D-particle background has important consequences for string matter propagating in such space-times, as it induces decoherence~\cite{emn}.
Indeed, let us first consider a $\sigma$-model propagating
in non-conformal backgrounds $\{ g^I \}$. The pertinent deformed world-sheet action reads:
\begin{equation}
S_\sigma = S^* + g^I \int_\Sigma V_I d^2\xi
\label{smodel}
\end{equation}
where $\Sigma$ is the world-sheet surface, $V_I$ are the appropriate vertex operators and $S^*$ is a conformal (fixed point) action. The above notation is schematic, but sufficient for our purposes.
In general, the notation $g^I$ denotes both the background-field species and the target-space argument of the fields, so the index $I$ runs over both continuous space-time coordinates and discrete (background species) values. For instance, if the deformation affects the space time metric, then $g^I = h_{MN}(y^P)$ where $h_{MN}$ is the deviation of the metric tensor from a fixed-point conformal background, and $M,N$ are target-space indices, while $y^P$ are target-space coordinates.

In our case, the recoil-velocity deformation is characterised by a coupling $g^I = u_i$, where $i$ is a target-space spatial index. In our model of foam, if the recoil of the D-particle is considered only on the D3 brane of fig.~\ref{fig:recoil}, then $i=1,2,3$. However, in general D-particles propagate in the bulk so, depending on the circumnstances and the model considered, $i$ can extend to the bulk spatial dimensions as well.
The corresponding vertex operator is the $D$-logarithmic operator (\ref{Depsilonop}). In the small-$\varepsilon $ limit, this is the dominant deformation, and for the purposes of this work we restrict our attention from now on to it.

In the above setting, if one considers a target-space quantity, such as the reduced density matrix $\rho_S$ of matter in an effective field theory limit, say, of strings propagating in such recoiling non-conformal $D$-particle backgrounds, then $\rho_S$ must be world-sheet renormalization group (WSRG) invariant, otherwise the world-sheet cut-off scale dependence would affect space-time quantities.
If we denote the appropriate WSRG evolution operator (i.e. the total derivative with respect to the WSRG scale ${\rm ln}\Lambda$) by $\frac{D}{D\,{\rm ln}\Lambda}$, then, the renormalizability  of the two-dimensional world-sheet $\sigma$-model theory, implies , for a \emph{fixed genus} world-sheet, the following:
\begin{equation}
\frac{D}{D\,{\rm ln}\Lambda} \rho_S = 0 = \frac{\partial}{\partial {\rm ln}\Lambda} \rho_S + \beta^I \frac{\partial}{\partial \,g^I}\rho_S + \dots = 0
\label{RGgroup2}
\end{equation}
where the $\dots $ denote other terms with implicit dependence on the cutoff, which we shall discuss below.  The set $\{ g^I \}$ denotes renormalized, ``running couplings/background fields''
in the $\sigma$-model, and as such $g^I = g^I ({\rm ln}\Lambda )$.
In the recoil case, under examination here, we have, as already mentioned,
\begin{equation}
 g^I \, \rightarrow \, u_i ~.
 \label{couplrecvel}
 \end{equation}
From the identification of the target-time  $t$ with the (logarithm) of the WSRG scale
${\rm ln}\Lambda $,
\begin{equation}
t \sim {\rm ln}\Lambda
\label{timelambda}
\end{equation}
as a result of (\ref{fsscaling}), (\ref{cdtrans}) and (\ref{scale2}) we then interpret, for this problem, Eq.~(\ref{RGgroup2}) as a \emph{time-evolution} equation in target-space~\cite{emn}.

We next notice that the WSRG derivatives of the couplings $g^I$, i.e. the WSRG $\beta$-functions $\beta^I \equiv \frac{d}{d\,t}g^I(t)$, are known in (perturbative) string theory to be derivable~\cite{osborn,emninfl,szabo}  from an off-shell target-space string-effective action ${\cal S}$,
\begin{equation}
\beta^I = G^{IJ} \frac{\delta {\cal S}}{\delta g^J}~,
\label{offshell}
\end{equation}
with $G^{IJ} \sim {\rm Lim}_{z \to 0} z^2{\overline z}^2 \langle V^I(z,{\bar z}), V^J(0.0) \rangle $ a Zamolodchikov metric in background $\{ g^I \}$ space~\cite{zam}, related to the short-distance behaviour  of the two-point correlation function on the world-sheet of the string ($z,{\overline z}$ denote world-sheet complexified variables, in a Euclidean world-sheet formalism, where the fixed-genus $\sigma$-model partition function is well defined).

In this sense, there is an effective Hamiltonian $H(g^I, p_I)$ in the target-space of strings, describing the dynamics of the background fields $g^I$, with $p_I$ the corresponding canonical momenta (in field theory space). The reduced density matrix $\rho_S$ depends in general on the ``phase-space'' variables $g^I, p_J$, hence the
$\dots $ in the WSRG equation (\ref{RGgroup2}) contain precisely this information about the $p_J$ dependence. The full result is~\cite{emn}:
\begin{equation}
\frac{D}{D t} \rho_S = 0 = {\dot \rho}_S + \beta^I \frac{\partial}{\partial \,g^I}\rho_S + {\dot p}_I \frac{\partial}{\partial \,p_I}\rho_S  = 0
\label{RGgroup}
\end{equation}
where the overdot denotes derivative with respect to time (\ref{timelambda}).

As already mentioned, upon the identification of time with a WSRG flow, the dynamical system $\{ g^I \}$ for a fixed genus world-sheet theory, satisfies the Helmholtz conditions so that the time-flow is viewed as a Hamiltonian flow~\cite{emn,emninfl,szabo}, with respect to the string effective Hamiltonian ${\cal H}(g^I, p_J)$. The non-conformal nature of the background fields $g^I$, though, affects the second Hamilton equation pertaining to the ``force'' ${\dot p_I}$ by terms proportional to $\beta^i$~\cite{emn}
\begin{equation}\label{hamilton}
\beta^I \equiv {\dot g}^I = \frac{\partial {\cal H}}{\partial p_I}~, \quad
{\dot p}_I = - \frac{\partial {\cal H}}{\partial g^I} + G_{IJ}\beta^J
\end{equation}
Classically, therefore, in target space the system will be characterised by a flow equation for the density matrix of the form:
\begin{equation}
{\dot \rho}_S = -\{\,\rho_S, \, {\cal H} \, \} - G_{IJ}\beta^J\frac{\partial}{\partial g^I}\rho_S
\end{equation}
with $\{ ~,~ \}$ denoting the appropriate Poisson bracket for the system.

Summation over genera on the world-sheet implies, as we have seen in the previous subsection \ref{subsec:genera}, a canonical quantization of the fields $g^I$, which can thus be replaced by appropriate quantum operators in the theory space
$g^I \to \widehat{g}^I$ while the Poisson brackets become quantum commutators $-i [ ~,~]$ and $\frac{\partial }{\partial p_I} \to -i[ {\widehat g}^I\,,~]$ following standard rules (in units of $\hbar =c=1$). In this way, the resulting evolution equation becomes  :
\begin{equation}\label{liouville}
\dot{\widehat{\rho}}_S = i[\,\widehat{\rho}_S, \, {\cal H} \,] + i: \widehat{G}_{IJ}
\widehat{\beta}^J [\widehat{g}^I\, , \, \widehat{\rho}_S]:
\end{equation}
where the hat notation denotes quantum operators and the $: \dots :$ denote the appropriate quantum ordering.

There is an important comment we wish to make here. In our approach, the target time flow has been identified with a WSRG flow. For consistency with the convergence of the world-sheet path intergral the target time $X^0$, appearing as a $\sigma$-model field, is necessarily Euclidean. To pass into Minkowskian signature one has to perform analytic continuation $X^0 \Rightarrow iX^0 $, which in turn implies that in a world-sheet path integral correlator  $\langle \dots \rangle  \Rightarrow i\langle \dots \rangle $, as a result of the measure of integration $\int DX^0 ...$. We should also take into account that, in stringy $\sigma$-models, the world-sheet Weyl (local conformal) invariance conditions are equivalent to equations of motion provided by a Lagrangian corresponding to (perturbative, on-shell) string S-matrix elements $\langle V_{I_1} \dots V_{I_n}\rangle_{\star} $, where $\langle \dots \rangle_{\star}$ indicates a path integral over the conformal invariant (fixed-point) $\sigma$-model action $S^\star$ (\emph{c.f.} (\ref{smodel})). This equivalence is expressed by means of the relation:
\begin{equation}\label{amplitudes}
G_{IJ}\beta^{J} = \langle V_{I} V_{I_1} \dots V_{I_n}\rangle_{\star} g^{I_1} \dots g^{I_n} 
\end{equation}
where the summation/integration is understood over repeated indices. Since the on-shell (Veneziano type) amplitudes $\langle V_{I} V_{I_1} \dots V_{I_n}\rangle_{\star}$ (which, by the way, are totally symmetric in their indices due to the well-known string dualities) involve path integration over the target time $\int DX^0$, with respect to a quadratic in $X^0$ $\sigma$-model conformal action (taken here over a flat target space-time initially),  we observe that,  upon analytic continuation of the time $\sigma$-model field $X^0 \to iX^0$, the term containing the Zamolodchikov metric $G_{IJ}^{({\rm Eucl,})}\beta^{J} \Rightarrow iG_{IJ}\beta^{J}$, and hence the Mikowskian-signature time evolution (\ref{liouville}) becomes finally:
\begin{equation}\label{liouville2}
\dot{\widehat{\rho}}_S = i[\,\widehat{\rho}_S, \, {\cal H} \,] - : \widehat{G}_{IJ}
\widehat{\beta}^J [\widehat{g}^I\, , \, \widehat{\rho}_S]:
\end{equation}
Notice that, as a result of the non-conformal nature of the background, the quantum version of the evolution equation for the density matrix of the string-matter subsystem includes an ``environmental decohering term''
\begin{equation}\label{decohliouv}
{\cal D}\widehat{\rho}_S \equiv -: \widehat{G}_{IJ}
\widehat{\beta}^J [\widehat{g}^I\, , \, \widehat{\rho}_S]:
\end{equation}
which cannot be cast in the form of a commutator with the effective hamiltonian ${\cal H}$, but the term is still linear in $\widehat{\rho}_S$.

The quantum ordering is then chosen so that the linear evolution equation (\ref{liouville2}) satisfies the axioms of positivity of the density matrix (whose diagonal elements are related to quantum probabilities), energy conservation on the average and probability conservation, which therefore implies a Lindblad formalism~\cite{lindblad}
(from now on we omit the hatted notation for quantum operators for brevity):
\begin{equation}\label{lindblad}
{\dot \rho} = i[\rho, {\cal H}] - \sum_{n} \left(\{ \rho, D_n^\dagger D_n\} - 2 D_n^\dagger \rho D_n^\dagger \right)
\end{equation}
where $D_n$ are the ``environmental'' operators inducing decoherence on the subsystem with reduced density matrix $\rho$.

For the case at hand, $g^I = u_i$ and $\beta^I = -\frac{\varepsilon^2}{2}u_i $, to lowest order in the (small) recoil velocities, and the only relevant components of the Zamolodchikov metric are~\cite{kogan}
 (c.f. Eq.~(\ref{canep3})) $G_{DD} \sim \frac{1}{\varepsilon^2} + \dots$, where the $\dots$ denote regular (in $\varepsilon \to 0$ limit) terms, which do not contribute to leading order in the small $\varepsilon$-expansion. In this way, we have a Lindblad system with environment operators
$D_i = u_i $, and decoherence of the double-commutator form:
\begin{equation}\label{urecoildec}
{\dot \rho}_S = i[\rho, {\cal H}] - \frac{1}{2}M_s [u_i, [u^i, \rho_S]]
\end{equation}
with $M_s$ the string scale (re-introduced here to make the units of energy of the decoherence coefficient apparent, and also to give concrete physical information on the order of magnitude of these terms, as we shall discuss below).

For completion we mention that formally the stochastic aspects of the quantum fluctuations of the recoil velocity, discussed in previous subsections, can be included formally by writing equation (\ref{urecoildec})
in a differential stochastic It\^o form~\cite{adler}, upon adding the appropriate It\^o stochastic differentials $dW_i$:
\begin{eqnarray}\label{stochliouville}
d\rho_S  & = &  i[\rho_S, {\cal H}]dt -\frac{1}{2}M_s [u_i, [u^i, \rho_S]] + \sqrt{M_s}[\rho_S, [\rho_S, u_i]]dW^i~, \nonumber \\ && dW_i dt   =   0,\quad dW_i dW^i =0~.
\end{eqnarray}
In our discussion below this will always be understood, although we shall not make explicit use of the It\^o calculus here.

The double commutator form of the Lindblad decoherence term due to the recoil in the D-particle foam case of interest here,
allows for a straightforward estimate in the case of a two-state quantum system, such as neutral-mesons or a dominant-two-flavour neutrino oscillation, which we shall concentrate our attention upon here.

The operator $\widehat{u_i}$ is not a simple, single particle operator, as $u_i = \frac{g_s}{M_s}\Delta k$ is proportional to the momentum transfer during the scattering of the matter string off the D0-brane. To simplify matters we can~\cite{mavromatos} represent this momentum transfer as a fraction of the initial momentum, in which case the relevant recoil-velocity operator can be written as:
\begin{equation}\label{rdef}
\widehat{u}_i = \frac{g_s}{M_s} r \widehat{k}~,
\end{equation}
In this parametrization it is possible to separate the statistical fluctuations of the recoil velocity, due to the population of D-particle defects in the foam (c.f. fig.~\ref{fig:recoil}) from the quantum fluctuations of a single recoil event, discussed in the previous two subsections. Both are assumed to be of a stochastic Gaussian nature.
For a single scattering of a matter string with a recoil defect, the quantum fluctuations, arising from a summation over world-sheet genera, as we have discussed in the previous two subsections, are denoted by $\langle \dots \rangle$ and are of Gaussian stochastic nature, about a zero mean $\langle \,u_i \rangle \, =0$ and variance $\sigma_0$ given by (\ref{recvelquantuflct}).

For the statistical fluctuations, the appropriate vacuum expectation values are denoted by $\langle\langle \dots \rangle\rangle $, and for the case at hand
are also assumed Gaussian
with zero average and non-trivial variance $\Delta$ for simplicity:
\begin{equation}\label{gausstat}
\langle \langle r \rangle \rangle=0, \quad  \langle \langle r^2 \rangle \rangle = \zeta^2 > 0~.
\end{equation}
However, the precise form of the distribution function of the statistical fluctuations of the D-particle recoil velocities is a feature of the specific microscopic model of string foam under consideration, and other cases, such as Cauchy-Lorentz distribution of populations of D-particle defects on our brane world, have also been studied~\cite{alexandre}. In our approach, though, as we do not want to consider Lorentz-violating effects of the foam on average,  we only use models with a zero mean of the recoil velocities of the space-time defects. One, however, may consider other models in which Lorentz symmetry is explicitly violated on average, in which case the effective low-energy field theory Hamiltonian would belong to the class of the Standard Model Extension (SME) considered in \cite{kostelecky}. We shall not discuss this latter case here.

 The statistical average over populations of D-particles
 can be taken at the level of the evolution equation (\ref{urecoildec}), under the parametrization (\ref{rdef}). For the Gaussian case (\ref{gausstat}), the result formally reads:
\begin{equation}\label{gaussevol}
\langle\langle \dot{\widehat{\rho}}_S \rangle\rangle = i \langle \langle [\widehat{\rho}_S, \widehat{{\cal H}}] \rangle\rangle \, - \, \frac{\zeta^2\, g_s^2}{2M_s} [\widehat{k_i}, [\widehat{k}^i, \rho_S]]
\end{equation}
where the operator nature, denoted by the hatted notation, indicates the result of genus summation on the world-sheet.

In the case of systems with two-mass eigenstates,
  with masses $m_i$, $i=1,2$ and momentum $\vec{k}$, $|\,\vec{k}, i\,\rangle $, it is straightforward to see from (\ref{gaussevol}) that in a center of mass frame, where one eigenstate has momentum +k and the other -k, only the non-diagonal matrix elements of the density matrix, corresponding to interference terms in the oscillation, exhibit non-trivial decoherence, leading to exponential damping of the oscillatory terms, with coefficients of the form:
\begin{equation}\label{dampinglind2}
 e^{-{\cal D} t}~, \quad {\cal D} \sim \zeta^2 \frac{g_s^2}{M_s} \langle \widehat{k}_i^2  \rangle
= \zeta^2 \frac{g_s^2}{M_s} |\vec k|^2
\end{equation}
where we took into account that $\vec{\widehat{k}}|\pm k, j\rangle = \pm \vec{k} |\pm k, j\rangle $, (no sum over $i,j=1,2$). In estimating the quantum fluctuations of $\widehat{k}^2$, we also took into account that
in our recoil case we assume quantum fluctuations about a zero $\langle u_i \rangle =0$ value, so that Lorentz symmetry is respected by the foam; hence the damping exponent (\ref{dampinglind2}) is directly proportional to the variance of the recoil velocity, $\sigma_0$, which has been calculated in detail in subsection \ref{subsec:langevin} (\ref{recvelquantuflct}). The statistical average over D-particle populations yields simply (c.f. (\ref{rdef})) the extra factor of $\zeta^2$.
This
leads to the following order of magnitude estimate of the Lindblad-decoherence  damping due to D-particle quantum and statistical fluctuations in our foam model:
\begin{equation}
   {\cal D} \sim \zeta^2 g_s^2 \frac{\overline{k}^2}{M_s}~,
   \label{dampinglind}
   \end{equation}
where $\overline{k}$ is a typical average momentum of the matter particle at hand (neutral Kaon or neutrino for the specific examples considered here), $g_s < 1$ is the (weak) string coupling and $M_s$ is the string mass scale. The reader should recall that the mass of the D-particle is $M_s/g_s$. The parameter $0 < \zeta^2 < 1 $ depends on the (statistical) details of the foam (type of distribution functions \emph{etc.}) and incorporates the probability of interaction of string matter with the D-particle defects. As such, it depends on the details of the microscopic foam models, such as density of D-particle defects on the brane world of fig.~\ref{fig:recoil} \emph{etc}. In the most optimistic case of having one D-particle per string-scale volume on the D3 brane world, $\zeta^2 ={\cal O}(1)$, which translates to an average momentum transfer during the flight of the matter particle of order of the initial incident energy. However, in realistic models one might have $\Delta \ll 1$, especially for dilute populations of D-particles. This may also depend on the cosmological era considered, since, depending on the model of foam, the distribution of D-particles in the bulk geometry of fig.~\ref{fig:recoil} may not be uniform at all.
Constraints of this parameter, therefore, may be imposed by studying in detail the cosmology of these models and comparing them against cosmological data on the Universe's energy budget.

Owing to the normal-ordering ambiguities of the decoherence term in (\ref{decohliouv}), a slightly more sophisticated  form of this term can be presented for the case of neutral Kaons (or neutrinos in that matter),
with CP and other properties of the system being taken properly into account. Indeed, our Lindblad decoherence (\ref{urecoildec}) can be cast in the paramertrization of ref.~\cite{ehns}, for neutral Kaon system, respecting the $\Delta S=\Delta Q$ rule, with $S$ the strangeness quantum number.
In such a case, the modified Lindblad evolution equation for
the respective density matrices of neutral kaon matter can be
parametrized as follows~\cite{ehns,lopez,huet}:
$$\partial_t \rho = i[\rho, H] + \delta\H \rho~,$$
where the Hamiltonian of the Kaon system is given by: {\small \begin{equation}
H_{\alpha\beta}=\left( \begin{array}{cccc}  - \Gamma
& -\coeff{1}{2}\delta \Gamma
& -{\rm Im} \Gamma _{12} & -{\rm Re}\Gamma _{12} \\
 - \coeff{1}{2}\delta \Gamma
  & -\Gamma & - 2{\rm Re}M_{12}&  -2{\rm Im} M_{12} \\
 - {\rm Im} \Gamma_{12} &  2{\rm Re}M_{12} & -\Gamma & -\delta M    \\
 -{\rm Re}\Gamma _{12} & -2{\rm Im} M_{12} & \delta M   & -\Gamma
\end{array}\right)
\label{kaonham}
\end{equation}
 and the Lindblad decoherence term has the form:
\begin{equation}
{\delta\H}_{\alpha\beta} =\left( \begin{array}{cccc}
 0  &  0 & 0 & 0 \\
 0  &  0 & 0 & 0 \\
 0  &  0 & -2\alpha  & -2\beta \\
 0  &  0 & -2\beta & -2\gamma \end{array}\right)~.
 \label{ehnsdec}
 \end{equation}
Positivity of $\rho$ requires: $\alpha, \gamma  > 0,\quad
\alpha\gamma>\beta^2$. Notice that $\alpha,\beta,\gamma$ violate
{\it both}  CPT, due to their decohering nature~\cite{wald}, and CP
symmetry, as they do not commute with the CP operator
$\widehat{CP}$~\cite{lopez}: $\widehat{CP} = \sigma_3 \cos\theta +
\sigma_2 \sin\theta$,\, $[\delta\H_{\alpha\beta}, \widehat{CP} ]
\ne 0$. In the context of our D-particle foam Lindblad decoherence model (\ref{urecoildec})
the statistical average over populations of D-particle defects, (\ref{gaussevol}),
has been properly taken prior to arriving at the parametrization (\ref{kaonham}), (\ref{ehnsdec}).

An important remark is now in order. As pointed out in
\cite{benatti}, although the above parametrization is sufficient for
a single-kaon state to have a positive definite density matrix (and
hence probabilities) this is \emph{not} true when one considers
the evolution of entangled kaon states (such as those encountered in $\phi$-factories~\cite{dafne}).
In this latter case,
complete positivity is guaranteed only if
the further conditions
\begin{equation}\label{cons}
\alpha = \gamma~ {\rm and} ~\beta = 0
\end{equation}
are imposed.
In this case, the Lindblad decoherence is described by a single parameter, $\gamma$, which leads to
exponential damping with time, of the form $e^{-\gamma \, t}$ for interference terms.

In the specific decoherence model (\ref{urecoildec}), (\ref{gaussevol}), considered here, which as we have seen above can be cast in this parameterization, respecting by construction complete positivity even on entangled states, we have estimated the order of this decoherence parameter as \emph{almost maximal}, see (\ref{dampinglind}),
in the sense that, up to a string-coupling factor $g_s < 1$, the rest
is the one expected from dimensional analysis $\overline{E}^2/M_{\rm QG}$,  since for the particle probes considered here (either neutral mesons or neutrinos) the energy scale is of the same order as the momentum scale and also, in the D-particle foam model, the quantum gravity scale is the D-particle mass, $M_s/g_s$.
The latter is usually assumed of the order of the four-dimensional Planck scale $10^{19}$ GeV, although one should keep an open mind, given the freedom one has in the values of $M_s, g_s$ in the modern version f string theory, in particular in low-$M_s$ string theories.

In the above discussion we have assumed that during the interaction of the string matter with the D-particle defect no ``flavour'' changes take place~\footnote{By ``flavour'' here we mean either the two CP eigenstates in the case of neutral mesons, or the two neutrino flavours in the neutrino case.}. However, this is not true in general. Indeed, the D-particles may be viewed as playing the r\^ole of space-time defects of brany black-hole type, appropriately compactified (say, D3-branes wrapped around 3 cycles, for instance see the type IIB construction of D-particle foam in ref.~\cite{li}). In such a way, the appropriate no-hair theorems of such black holes allows ``flavour''  non-conservation during the interaction, and therefore oscillations. In other words, in the simple models of D-foam depicted in fig.~\ref{fig:recoil}, the re-emitted string state after capture
by the defect may be of different flavour than the incident one. This implies that in our effective representation (\ref{rdef}) of the recoil velocity in terms of fractions of the momentum operator the function $r$ is no longer a scalar function, but assumes a matrix structure in flavour space. In the simplified case of two flavours, examined here,
we thus have the replacement (at an effective low-energy field-theory level):
\begin{equation}
               r \to \underline{\underline{r}} = r_0 I_{2\times 2} + \sum_{i=1}^{3} r_i \sigma_i \equiv r_\mu \sigma_\mu ~, \quad <r_\mu> =0,~ <r_\mu r_\nu>=\delta_{\mu\nu} \Delta_\mu   \label{mutliflavour}
\end{equation}
where $\sigma_i$ are $2 \,\times\, 2 $ Pauli matrices, and $I_{2\times 2}$ is the $2 \,\times\, 2 $ identity matrix. Inclusion of such flavour changes does not affect the order of magnitude estimate of the Lindblad-decoherence effects (\ref{dampinglind}).

Perhaps it should be remarked at this stage that the current model of foam is different from the \emph{energy-driven decoherence} model of Adler and Horwitz~\cite{adler}, presented as a rather generic Lindblad-type model of quantum-gravity
foam in two-level quantum systems, which leads to much more suppressed decoherence effects. In \cite{adler} it was argued that,
 under the additional assumption of entropy increase and exact energy conservation in the system, hermiticity of the Lindblad environmental operators, $D_n$, and commutativity with the Hamiltonian $\widehat{{\cal H}}$ follow, which in the case of a two-level system imply
 either $D_n \propto I_{2 \times 2 }$ or $D_n \propto \widehat{{\cal H}}$. In the latter case, the double commutator Lindblad form for the (single) operator $D$, assumed for simplicity, implies a much more suppressed decoherence-damping exponent,  proportional to the energy difference between the two eigenstates
 \begin{equation}\label{adlermodel}
     {\cal D} \sim \frac{(\Delta E)^2}{M_{\rm QG}}~,
\end{equation}
where $M_{\rm QG}$ is a typical quantum gravity scale, depending on the microscopic model. For the neutral kaon or neutrino oscillating systems this term is of order
$\frac{(\Delta m^2)^2}{E^2\, M_{\rm QG}}$, where $\Delta m^2 \equiv m_1^2 - m_2^2 $ the appropriate difference between the squared of the masses of the two mass eigenstates, which is assumed much smaller (as is the physical case) than then typical energies $E$ of the matter probes.
Indeed, for neutral Kaons, for instance, $\Delta m  \sim 10^{-15}$ GeV, while the typical energies in a $\phi$-factory~\cite{dafne} or a single-beam Kaon experiment, such as CPLEAR~\cite{cplear}, are of order of GeV. For neutrinos, $\Delta m^2 $ is at most $10^{-3}$ eV$^2$, with typical energies higher than MeV for experiments of oscillations of interest to probe quantum gravity effects~\cite{neutrino}. We thus see that the Adler-Horwitz decoherence model (\ref{adlermodel}), seems, at least presently, beyond experimental reach.

In contrast to the estimates implied by  this model, our D-particle foam model leads to much stronger decoherence effects (\ref{dampinglind}), provided the parameter $\zeta^2$ is not too small, i.e. the population of D-particles is not too dilute. This is mainly due to the fact that, in our case, due to momentum transfer between string matter and the D-particle, there is energy conservation only on average, and thus the environmental operators are proportional to the momentum transfer, and hence have a much more complicated structure than in the simple model of \cite{adler}. The combination
\begin{equation}
M_{\rm QG~eff} \equiv \frac{M_s}{\zeta^2 \, g_s^2}~,
\label{qgeff}
\end{equation}
in the D-particle foam model plays the r\^ole
of an ``effective quantum-gravity scale'' that
can be constrained by experiment, in particular neutrino oscillations~\cite{neutrino}, which seem to offer at present the highest possible sensitivity on bounding quantum-gravity-decoherence-induced damping effects in particle oscillations. In fact, the effective quantum-gravity scale from neutrino oscillations can be currently pushed as much as $10^{27}$ GeV, which on account of (\ref{qgeff}), for natural values of $M_s/g_s \sim 10^{19}$ GeV, yields the upper bound $\zeta^2 g_s \le  10^{-8}$.
We remind the reader that the statistical variance $\zeta^2$ of the recoil velocity of the D-particle populations is a quantity independent of the string scale, that depends on statistical microscopic properties of the D-particle population, which may be cosmological-era dependent (c.f. fig.~\ref{fig:recoil}).
Therefore the present-era bounds may not be translated trivially into the early Universe, where the density of D-particles in the foam might have been much higher.

\subsection{D-particle-recoil-induced Finsler-metric distortions and Space-Time Non Commutativity\label{sec:lcftim}}

In addition to the Lindblad-type decoherence, implied by stochastic quantum fluctuations of the D-particle recoil velocity in the foam, there are also induced modifications of the target-space metric `felt' by the open string describing matter excitations in interaction with the foam. There is a straightforward way to see this, by drawing an analogy of the existence of the recoil velocity term, for times $X^0 > 0$ well after the
string-splitting/capture event, with an electric background field~\cite{seibergwitten,sussk1}.

Indeed, let one consider the boundary recoil/capture operator $\mathcal{V}_{\rm{D}}^{imp}$ (\ref{fullrec}) in the Dirichlet picture, written as a total derivative over the bulk of the world-sheet by means of the two-dimensional version of Stokes theorem (omitting from now on the explicit summation over repeated $i$-index, which is understood to be over the spatial indices of the D3-brane world):
\begin{eqnarray}\label{stokes}
&& \mathcal{V}_{\rm{D}}^{imp}=\frac{1}{2\pi\alpha '}
\int_{D}d^{2}z\,\epsilon_{\alpha\beta} \partial^\beta
\left(  \left[  u_{i}X^{0}\right]  \Theta\left(  X^{0}\right)  \partial^{\alpha}X^{i}\right) = \nonumber \\
&& \frac{1}{4\pi\alpha '}\int_{D}d^{2}z\, (2u_{i})\,\epsilon_{\alpha\beta}
 \partial^{\beta
}X^{0} \Bigg[\Theta_\varepsilon \left(X^{0}\right) + X^0 \delta_\varepsilon \left(  X^{0}\right) \Bigg] \partial
^{\alpha}X^{i}
\end{eqnarray}
where $\delta_\varepsilon (X^0)$ is an $\varepsilon$-regularised $\delta$-function.
This is equivalent to a deformation describing an open string propagating in an antisymmetric  $B_{\mu\nu}$-background corresponding to an external constant in target-space ``electric'' field,
\begin{equation}
B_{0i}\sim u_i ~, \quad B_{ij}=0~,
\label{constelectric}
\end{equation}
where the $X^0\delta (X^0)$ terms in the argument of the electric field yield vanishing contributions in the large time limit $\varepsilon \to 0$, and hence are ignored from now on~\footnote{We remark for completeness that, upon a T-duality canonical transformation of the coordinates~\cite{tdual}, the presence of the B-field leads to mixed-type boundary conditions for open strings on the boundary $\partial \mathcal{D}$  of world-sheet surfaces with the topology of a disc:
\begin{equation}
      g_{\mu\nu}\partial_n X^\nu + B_{\mu\nu}\partial_\tau X^\nu |_{\partial \mathcal{D}} = 0~,
\label{bc}
\end{equation}
with $B$ given by (\ref{constelectric}). Absence of a recoil-velocity $u_i$-field leads to the usual Neumann boundary conditions, while the limit where $G_{\mu\nu} \to 0$, with $u_i \ne 0$, leads to Dirichlet boundary conditions.}.

Considering commutation relations among the coordinates of the first quantised $\sigma$-model in the above background, one also obtains a non-commutative space-time relation~\cite{sussk1}~\footnote{In contrast, in the case of strings in a constant magnetic field~\cite{seibergwitten} (corresponding to B-fields of the form $B_{ij} \ne 0$, $B_{0i~}=0$), the non commutativity is only between spatial target-space coordinates}. The pertinent non commutativity refers to spatial coordinates along the direction of the electric field, and is expressed in the form
\begin{equation}
[ X^1, t ] = i \theta^{10} ~, \qquad \theta^{01} (= - \theta^{10}) \equiv \theta =  \frac{1}{u_{\rm c}}\frac{\tilde u}{1 - \tilde{u}^2}
\label{stnc}
\end{equation}
where $t$ is the target time, and we assume for simplicity and concreteness recoil along the spatial $X^1$ direction. The quantity $\tilde{u}_i \equiv \frac{u_i}{u_{\rm c}}$ and  $u_{\rm c} = \frac{1}{2\pi \alpha '}$ is the Born-Infeld \emph{critical} field.
Notice that the presence of the critical ``electric'' field is associated with a singularity of both the effective metric and the non commutativity parameter, while, as we shall discuss below (\ref{effstringcoupl}) there is also an effective string coupling, which  vanishes in that limit. This reflects the \emph{destabilization of the vacuum} when the ``electric'' field intensity approaches the \emph{critical value}, which was noted in \cite{burgess}. Since in our D-particle foam case, the r\^ole of the `electric' field is played by the recoil velocity of the
D-particle defect, the critical field corresponds to the relativistic speed of light, in accordance with special relativistic kinematics, which is respected in string theory, by construction.

The space-time uncertainty relations (\ref{stnc}) are consistent with the corresponding space-time string uncertainty principle~\cite{yoneya}
\begin{equation}
   \Delta X \Delta t \ge \alpha '
\label{stringyunc}
\end{equation}
As discussed in detail in refs.~\cite{sussk1,seibergwitten},
 there is also an induced open-string \emph{effective target-space-time metric}.
To find it,
one should consider the world-sheet propagator on the disc $\langle X^\mu(z,{\overline z})X^\nu(0,0)\rangle$, with the boundary conditions (\ref{bc}).
Upon using a conformal mapping of the disc onto the upper half plane
with the real axis (parametrised by $\tau \in R$) as its boundary~\cite{seibergwitten},
one then obtains:
\begin{equation}
     \langle X^\mu(\tau)X^\nu(0)\rangle = -\alpha ' g^{\mu\nu}_{\rm open,~electric}{\rm ln}\tau^2 + i\frac{\theta^{\mu\nu}}{2}\epsilon(\tau)
\label{propdisc}
\end{equation}
with the non-commutative parameters  $\theta^{\mu\nu}$ given by by (\ref{stnc}),
and the effective open-string metric, due to the presence of the recoil-velocity field $\vec{u}$, whose direction breaks target-space Lorentz invariance, by:
\begin{eqnarray}
           g_{\mu\nu}^{\rm open,electric} &=& \left(1 - {\tilde u}_i^2\right)\eta_{\mu\nu}~, \qquad \mu,\nu = 0,1 \nonumber \\
           g_{\mu\nu}^{\rm open,electric} &=& \eta_{\mu\nu}~, \mu,\nu ={\rm all~other~values}~,
\label{opsmetric}
\end{eqnarray}
where, for concreteness and simplicity, we consider a frame of reference where the matter particle
has momentum only across the spatial direction $X^1$, \emph{i.e.} $0 \ne k_1 \equiv k \parallel u_1~, k_2=k_3 =0$.
Moreover, there is a modified effective string coupling~\cite{seibergwitten,sussk1}:
\begin{equation}
   g_s^{\rm eff} = g_s \left(1 - \tilde{u}^2\right)^{1/2}
\label{effstringcoupl}
\end{equation}
The fact that the metric in our recoil case depends on momentum transfer variables, implies that D-particle recoil induces Finsler-type metrics~\cite{finsler}, \emph{i.e}. metric functions that depend on phase-space coordinates, that is space-time and momentum coordinates. We mention here that such metrics have been suggested in the context of a T-dual Neumann picture~\cite{szabo} of the D-particle recoil process in refs.~\cite{Dfoam} (we note that T-duality is a canonical transformation in the $\sigma$-model path integral~\cite{tdual}).

Stochastic quantum fluctuations of the recoil velocity $u_i$, therefore, due to both statistical (D-particle population) and genuine quantum (genus summation) effects imply corresponding fluctuations in the induced space-time metric (\ref{opsmetric}) and, through them, they  affect the propagation of matter strings on such fluctuating geometries.
Such metric fluctuations will modify the dispersion relations of the matter probe,
and hence will be associated with  the Hamiltonian commutator term in the density-matrix evolution equation (\ref{urecoildec}). This term will contribute to a phase in the density matrix, and hence such metric fluctuations will not contribute to damping but affect the oscillation period. For the induced metric (\ref{opsmetric}), we obtain the modified dispersion relation:
\begin{equation}\label{disper}
            \omega_i^2 = k^2 + \frac{m_i^2}{1 - u_1^2} \simeq k^2 + m_i^2 + m_i^2 u_1^2 + \dots ~, \quad i=1,2~,
\end{equation}
to leading order in the small recoil velocities $|u_1| \ll 1$, where we restrict our attention throughout this work.

For concreteness, let us consider the case of high energy matter particles, for which $k \gg m_i$. In such a case, it is straightforward to see from (\ref{urecoildec}) that the shift in the oscillation period, due to the modified dispersion relation (\ref{disper}) in our D-particle foam model, is of order:
\begin{equation}\label{shift}
{\rm Shift~in~argument~of~trigonometric~oscillatory~terms}: \, g_s^2\frac{m_2^2 - m_1^2}{2\,M_s^2}\zeta^2 k t
\end{equation}
where $t$ is the time. Compared to the standard oscillation term $(m_2^2-m_1^2)t/2k$, this is negligible for D-particles in the (natural) range of masses $M_s/g_s \sim 10^{19}$ GeV, and a wide range of energies/momenta  $k$ of matter probes, even as high as those of high-energy cosmic rays $k \sim 10^{20}$ eV.

For completeness,  we also mention at this stage that, when we consider the physically interesting case of ``flavoured matter'' with mixing, such as neutrinos, then the mass eigenstates are different from the physically observed
flavour eigenstates, which are related to the former by
$|\phi_{\alpha}\rangle= \sum_{i} U_{\alpha i} |f_{i}\rangle$, with Greek indices $\alpha$ denoting flavour, and Latin indices denoting mass eigenstates.
For instance, for the simple but instructive case of two flavours, with mixing angle $\theta$, the matrix $U$ is given by the $2 \times 2 $ unitary matrix:
$$ U = \left(\begin{array}{cc} {\rm cos}\theta \quad &{\rm sin}\theta \\ -{\rm sin}\theta
\quad & {\rm cos}\theta \end{array}\right)\ ~.$$ More complicated expressions, with more than one mixing angles are considered for the physically relevant case of three light neutrino flavours~\cite{neutrino}.
Formally, the oscillation probability between the flavour eigenstates $\alpha \to \beta$ is expressed in terms of the flavour density matrix
$$ {\cal P}_{\alpha \to \beta} ={\rm Tr}[\rho_\beta (t) \rho_\alpha(0)]~,
\quad \rho_{\alpha}(0) \equiv |\,\alpha\rangle\,\langle \alpha \,|~,$$
and the time-dependent flavour density matrix $\rho_\beta(t)$ is found by solving appropriately the Lindblad-evolution equation, including mixing as described above.
One also takes the statistical average over the population of D-particle defects in the foam~\cite{alexandre}. The flavour mixing parameters (angle) appear as multiplicative trigonometric coefficients to the formulae for the induced Lindblad damping exponents in this case, and hence for mixing angles in the physically relevant range (for, say, neutrinos), such terms are of order one and do not affect significantly the above-mentioned order of magnitude estimates (\ref{dampinglind}) of decoherence in the model.

To recapitulate, in this string-inspired model of D-particle foam, the decoherence, which is of Lindblad type, owes its existence to the environmental terms of (\ref{urecoildec}), due to the non-conformal $\sigma$-model D-particle recoil deformations (\ref{fullrec}).
A final, but important, aspect of this type of string foam is the induced intrinsic CPT violation, as a result of this decoherence, which we now come to discuss, paying particular attention to its potential experimental consequences.

\section{Intrinsic CPT Violation\label{sec:3}}

\subsection{Microscopic Time Arrow due to Decoherence}

The presence of decoherence in the effective low-energy string-inspired field theory limit, describing propagation of low-energy string matter in D-particle foam, implies a strong form of ``violation'' of CPT operator in the sense of \cite{wald}.
Indeed, according to that work, the evolution of initially pure quantum states to mixed ones, due to the decoherent evolution of an open quantum relativistic system, will result in an ill-defined CPT operator.
The proof is essentially based~\cite{wald} on initially making the assumption of a well-defined unitary~\footnote{The CPT operator $\theta$ acting on state vectors is anti-unitary, due to the anti-unitary nature of the time-reversal operator, however when one considers the
action of the CPT operator $\Theta$ on density matrices $|x\rangle \langle y|$ then it is unitary, as it is essentially a product of $\theta^\dagger \theta $~.} CPT operator, $\Theta$, and then arriving at an inconsistency, thus proving the initial assumption wrong.
Consider the CPT operator $\Theta$,
acting on asymptotic (`in' and `out') density matrices,
\begin{equation}\label{thetacpt}
\Theta \rho_{\rm in} = {\overline \rho}_{\rm out}~, \quad \Theta^\dagger = \Theta^{-1}
\end{equation}
 One should take into account that in decoherent quantum systems the asymptotic density matrices are related by the (linear) super-scattering operator of Hawking~\cite{hawking}:
\begin{equation}\label{dollar}
\rho_{\rm out} = \$ \rho_{\rm in}~, \qquad
\overline{\rho}_{\rm out} = \$\overline{\rho}_{\rm out}
\end{equation}
In decoherent situations, the factorizability property of the
super-scattering matrix \$ in terms of oridnary scattering S-matrix breaks down, \$ $\ne SS^\dagger $,
where $S=e^{-H t}$, with $H$ the Hamiltonian. This property is equivalent to the breakdown of the local effective lagrangian formalism, which relies on well-defined perturbative scattering matrices.
It is important to notice that, as a result of ``loss of information'' for the asymptotic low-energy observer, who cannot measure degrees of freedom of the system trapped inside either microscopic quantum-fluctuating horizons in  generic models of space-time foam~\cite{wheeler}, or, in the particular case of D-particle foam~\cite{Dfoam}, associated with space-time distortions in the neighborhood of the D-particle defect as a result of the capture/splitting process of fig.~\ref{fig:recoil}, the $\$ $-matrix has~\cite{hawking,wald} \emph{no inverse.}

From (\ref{thetacpt}) and (\ref{dollar}), we readily obtain:
\begin{eqnarray}
{\overline \rho}_{\rm in} = \Theta \rho_{\rm out} = \Theta \, \$ \, \rho_{\rm in} =
\Theta \, \$ \, \Theta^{-1} \Theta \rho_{\rm in} =
 \Theta \, \$ \, \Theta^\dagger {\overline \rho}_{\rm out} =
\Theta \, \$ \, \Theta^{\dagger} \, \$ \, {\overline \rho}_{\rm in}
\end{eqnarray}
from which it follows that $ \Theta \$ \Theta^\dagger $ is the inverse of $\$ $.
This contradicts the above-mentioned property of $\$ $ of having no-inverse, and implies that the assumption on the existence of a well-defined unitary CPT operator $\Theta $, acting on density matrices, was \emph{false}.

This is really a breakdown of a \emph{Microscopic Time Irreversibility}~\cite{wald}, due to
the afore-mentioned loss of information in the low-energy subsystem due to decoherence.
This time irreversibility is, in principle, unrelated to CP properties, and this implies a breakdown of the entire CPT symmetry, in the sense of the afore-mentioned ill-defined nature of the quantum generator of CPT symmetry (\emph{intrinsic} CPT Violation).
The result should come to no surprise, since decoherence violates one of the important assumptions on CPT theorem~\cite{cpt}, that of unitarity. In our model of foam, of course
there is also a second assumption that is violated locally in space time, that of Lorentz symmetry~\cite{cpt,greenberg}, due to the direction of the recoil velocity of the D-particle defect. Even if, Lorentz symmetry is preserved on the average, due to $\langle u_i \rangle = 0$, nevertheless, fluctuations of the vector $u_i $ are non trivial, $\langle u_i^2 \rangle \ne 0$.

Notice, though, that the violation of CPT in our model is primarily related to the ill-defined nature of the CPT generator as a result of the non-conformal (on the world-sheet) nature of
the recoiling D-particle background, over which the matter string propagates.
As explained above, it is the deviation from conformal invariance that leads to extra decoherence terms in the respective evolution equation (\ref{urecoildec}) of the reduced density matrix of matter in such an ``environment''. Therefore, although - as we have seen in the previous subsection (c.f. (\ref{stnc})) - the D-particle recoil induces space-time \emph{non-commutativity}, it is \emph{not} the latter that implies microscopic Time irreversibility, and thus~\cite{wald} \emph{intrinsic} CPT Violation, but rather the decoherence due to the non-conformal nature of the recoiling (changing also with time, due to the impulse operators $\Theta(X^0)$) background over which the string propagates.
That it is not the space-time non commutativity that causes the CPT Violation is also in agreement with the analogy of the D-particle recoil with that of a constant electric field~\cite{seibergwitten}. It is known~\cite{carrol}, that such a non-commutative field-theory case leads to an effective low-energy local field theory lagrangian of the SME type~\cite{kostelecky}, in the continuum limit, with Lorentz but \emph{not} CPT Violating terms (the Lorentz violation is due to the direction of the electric field, whose r\^ole is played here by the direction of the recoil velocity).

It should be stressed, that in our approach, the ill-defined nature of the CPT operator is only \emph{perturbative}, in the sense that the anti-particle state exists.
It is a feature of the effective low-energy limit of string theory, where measurements are performed by a low-energy observer, using local (super)scattering matrices, which cannot detect the quantum fluctuating distortions of space time, as a result of the recoiling defects during their interaction with low-energy string matter. The full string theory is of course a well-defined theory of quantum gravity, and should be characterised by some version of CPT theorem although, I must admit that a formally constructed CPT operator, that preserves the symmetry at a fully non-perturbative string-theory level is way beyond the author's understanding at present.
For attempts to argue that a non-perturbative version of string/M-theory might conserve CPT exactly,
I refer the reader to some recent articles in the literature~\cite{dine}. However, in view of the ``foamy '' constructions mentioned in the current article and past works on the subject mentioned in its references,
I do not share the opinion that this issue can be settled easily, and in this sense experimental searches for possible CPT violations in the context at least of low-energy limits of quantum gravity, should be pursued.
This is a topic I turn to now, in an attempt to discuss rather unique, ``smoking-gun type'' effects
in some experimental situations, where decoherence-induced CPT violation, if true, can be tested in a rather unambiguous way.

\subsection{Intrinsic CPT Violation and Entangled Neutral-Meson States}

\subsubsection{The $\omega$-effect}

We now come to a description of an entirely novel effect~\cite{bmp}
of CPT Violation (CPTV) due to the ill-defined nature of the CPT
operator, which is rather \emph{exclusive} to neutral-meson factories, for
reasons explained below. The effect, termed $\omega$-effect~\cite{bmp},
is associated with appropriate modifications of the Einstein-Podolsky-Rosen (EPR)
correlators of entangled neutral meson states in a meson factory,
These
effects are qualitatively similar
for Kaon~\cite{bmp,bmpw} and $B$-meson factories~\cite{bomega}, but
in kaon factories there is a particularly
good channel, that of both correlated kaons decaying to $\pi^+\pi^-$.
In that channel the sensitivity of the $\omega$-effect increases because
the complex parameter $\omega$, parametrizing the relevant
EPR modifications~\cite{bmp}, appears in the particular
combination $|\omega|/|\eta_{+-}|$, with
$|\eta_{+-}| \sim 10^{-3}$. In the case of  $B$-meson factories
one should focus instead on the ``same-sign'' di-lepton
channel~\cite{bomega}, where high statistics occurs.

We commence
our discussion by briefly reminding the reader of
EPR particle correlations.
The EPR effect was originally proposed as a {\it paradox}, testing the
foundations of Quantum Theory. There was the question whether
quantum correlations between spatially separated events implied
instant transport of information that would contradict special relativity.
It was eventually realized that no super-luminal propagation was
actually involved in the EPR phenomenon, and thus there was no
conflict with relativity.

\begin{figure}[ht]
\begin{center}
  \includegraphics[width=0.4\textwidth]{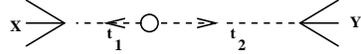}
\end{center}
\caption{Schematic representation of the decay of a $\phi$-meson at
rest (for definiteness) into pairs of entangled neutral kaons, which
eventually decay on the two sides of the detector.}
\label{epr}\end{figure}

The EPR effect has been confirmed experimentally, \emph{e.g.}, in meson
factories: (i) a pair of particles can be created in a definite
quantum state, (ii) move apart and, (iii) eventually decay when they
are widely (spatially) separated (see Fig.~\ref{epr} for a schematic
representation of an EPR effect in a meson factory). Upon making a
measurement on one side of the detector and identifying the decay products,
we \emph{infer} the type of products appearing on the
other side; this is essentially the EPR correlation phenomenon.
It does \emph{not} involve any \emph{simultaneous measurement} on
both sides, and hence there is no contradiction with special
relativity. As emphasized by Lipkin~\cite{lipkin}, the EPR
correlations between different decay modes should be taken into
account when interpreting any experiment.

In the case of $\phi$ factories
it was \emph{claimed }~\cite{dunietz} that
due to EPR correlations,  \emph{irrespective} of CP, and CPT
violation, the \emph{final state} in $\phi$ decays: $e^+ e^-
\Rightarrow \phi \Rightarrow K_S K_L $ always contains $K_LK_S$
products.
This is a direct consequence
of imposing the requirement of {\it Bose statistics}
on the state $K^0 {\overline K}^0$ (to which the $\phi$ decays);
this, in turn, implies that the physical neutral meson-antimeson
state must be {\it symmetric} under C${\cal P}$, with C the charge
conjugation and ${\cal P}$ the operator that permutes the spatial
coordinates. Assuming {\it conservation} of angular momentum, and a
proper existence of the {\it antiparticle state} (denoted by a bar),
one observes that: for $K^0{\overline K}^0$ states which are
C-conjugates with C$=(-1)^\ell$ (with $\ell$ the angular momentum
quantum number), the system has to be an eigenstate of the
permutation operator ${\cal P}$ with eigenvalue $(-1)^\ell$. Thus,
for $\ell =1$: C$=-$ $\rightarrow {\cal P}=-$.  Bose statistics
ensures that for $\ell = 1$ the state of two \emph{identical} bosons
is \emph{forbidden}. Hence, the  initial entangled state:
{\begin{eqnarray*} &&|i> = \frac{1}{\sqrt{2}}\left(|K^0({\vec
k}),{\overline K}^0(-{\vec k})>
- |{\overline K}^0({\vec k}),{K}^0(-{\vec k})>\right)  \nonumber \\
&& = {\cal N} \left(|K_S({\vec k}),K_L(-{\vec k})> - |K_L({\vec
k}),K_S(-{\vec k})> \right)\nonumber
\end{eqnarray*}}with the normalization factor ${\cal N}=\frac{\sqrt{(1
+ |\epsilon_1|^2) (1 + |\epsilon_2|^2
)}}{\sqrt{2}(1-\epsilon_1\epsilon_2)} \simeq \frac{1 +
|\epsilon^2|}{\sqrt{2}(1 - \epsilon^2)}$, and
$K_S=\frac{1}{\sqrt{1 + |\epsilon_1^2|}}\left(|K_+> + \epsilon_1
|K_->\right)$, $K_L=\frac{1}{\sqrt{1 + |\epsilon_2^2|}}\left(|K_->
+ \epsilon_2 |K_+>\right)$, where $\epsilon_1, \epsilon_2$ are
complex parameters, such that $\epsilon \equiv \epsilon_1 +
\epsilon_2$ denotes the CP- \& T-violating parameter, whilst $\delta
\equiv \epsilon_1 - \epsilon_2$  parametrizes the CPT \& CP violation
within quantum mechanics~\cite{fide}, as discussed previously.
The $K^0 \leftrightarrow {\overline K}^0$ or $K_S
\leftrightarrow K_L$ correlations are apparent after evolution, at
any time $t > 0$ (with $t=0$ taken as the moment of the $\phi$ decay).

In the above considerations there is an implicit assumption,
which was noted in \cite{bmp}. The above arguments are valid
independently of CPTV but only under the provision that such violation occurs
within quantum mechanics, \emph{e.g.}, due to spontaneous Lorentz
violation~\cite{kostelecky}, where the CPT operator is well defined.

If, however, CPT is \emph{intrinsically} violated, in the sense of being ill-defined
as a result of
decoherence in space-time foam models~\cite{wald}, the
concept of the ``antiparticle'' may be \emph{modified} perturbatively! The
perturbative modification of the properties of the antiparticle is
important, since the antiparticle state is a physical state which
exists, despite the ill-definition of the CPT operator. However, the
antiparticle Hilbert space will have components that are
\emph{independent} of the particle Hilbert
space.

In such a case,
the neutral mesons $K^0$ and ${\overline K}^0$ should \emph{no
longer} be treated as \emph{indistinguishable particles}. As a
consequence~\cite{bmp}, the initial entangled state in $\phi$
factories $|i>$, after the $\phi$-meson decay, will acquire a component
with opposite permutation (${\cal P}$) symmetry:
{ \begin{eqnarray}\label{initialomega} |i> &=& \frac{1}{\sqrt{2}}\left(|K_0({\vec
k}),{\overline K}_0(-{\vec k})>
- |{\overline K}_0({\vec k}),K_0(-{\vec k})> \right)\nonumber \\
&+&  \frac{\omega}{2} \left(|K_0({\vec k}), {\overline K}_0(-{\vec k})> + |{\overline K}_0({\vec
k}),K_0(-{\vec k})> \right)  \bigg]  \nonumber \\
& = & \bigg[ {\cal N} \left(|K_S({\vec
k}),K_L(-{\vec k})>
- |K_L({\vec k}),K_S(-{\vec k})> \right)\nonumber \\
&+&  \omega \left(|K_S({\vec k}), K_S(-{\vec k})> - |K_L({\vec
k}),K_L(-{\vec k})> \right)  \bigg]~,
\end{eqnarray}}where ${\cal N}$ is an appropriate normalization factor,
and $\omega = |\omega |e^{i\Omega}$ is a complex parameter,
parametrizing the intrinsic CPTV modifications of the EPR
correlations. Notice that, as a result of the $\omega$-terms, there
exist, in the two-kaon state,
$K_SK_S$ or $K_LK_L$ combinations,
which
entail important effects to the various decay channels. Due to this
effect, termed the $\omega$-effect by the authors of \cite{bmp},
there is \emph{contamination} of ${\cal P}$(odd) state with ${\cal P}$({\rm even})
terms. The $\omega$-parameter controls the amount of contamination
of the final ${\cal P}$(odd) state by the ``wrong'' (${\cal P}$(even)) symmetry state.
A time evolution of the $\omega$-terms, even in a purely unitary Hamiltonian evolution,
will lead~\cite{bmp,bmpw,bomega} to observable differences in the final states, as compared with the CPT conserving case, that can be tested experimentally in principle, as we shall describe briefly in subsection \ref{sec:omegaobs} below, and in fact constitute, if observed, rather ``smoking-gun'' evidence of this type of decoherence-induced CPT Violation. Before doing this, however, it is essential to describe the $\omega$-like effects that arise in the specific model of D-particle foam, and estimate the order of magnitude of such effects.

\subsubsection{Searching for $\omega$-like effects in D-particle Foam}

In the context of our D-particle foam model, the induced decoherence is responsible for the generation
of $\omega$-like terms $K_SK_S$ or $K_LK_L$ (or better, appropriate combinations of $K^0 K^0 $ and ${\overline K}^0 {\overline K}^0$
terms), due to the decoherent evolution in the space-time
foam (\ref{urecoildec}). In fact, as discussed in \cite{bmpw}, in the parameterization (\ref{ehnsdec}), (\ref{cons}), extended appropriately to entangled Kaon states~\cite{huet}, $\omega$-like terms $K_SK_S$ or $K_LK_L$, generated by decoherent time evolution, appear with coefficients~\cite{bmpw}
\begin{equation}
-\frac{2\gamma}{\Delta \Gamma} \rho_S \otimes \rho_S~, \quad  \frac{2\gamma}{\Delta \Gamma} \rho_L \otimes \rho_L~
\label{gammadec}
\end{equation}
with $\rho_{S(L)}$ appropriate density matrices for Short(S)- and Long (L) -lived Kaons respectively, and $\Delta \Gamma = \Gamma_S - \Gamma_L \sim 10^{-15} $~GeV.
In our D-particle foam model, the Lindblad decoherence coefficient $\gamma$ is of the order of ${\cal D}$ in (\ref{dampinglind}), \emph{i.e.}
\begin{equation}
\gamma \sim g_s^2 \zeta^2 \frac{\overline{k}^2}{M_s}~,
   \label{dampinglindgamma}
   \end{equation}
with $\overline{k}$ a typical average momentum of the matter particle. We shall discuss the experimental prospects for detecting such effects in subsection \ref{sec:exp}.

A natural question to ask at this stage is whether $\omega$-like terms, of the form (\ref{initialomega}), could appear in this model.
If the initial decay of the $\phi$-meson takes place in the presence of a D-particle defect, which is a natural assumption to make, then there would also be $\omega$-like decoherent terms in the initial state of the two kaons, of the form (\ref{initialomega}).  Indeed, the presence of a D-particle in the initial
entangled state of two neutral mesons, after the $\phi$-meson decay, implies perturbations in the Hamiltonian of the system, due to the induced metric fluctuations (\ref{opsmetric}), which, in the small $u_i$ limit, yield modified dispersion relations for the matter probes of the form (\ref{disper}).
Solving for the energy $\omega$, by taking the square root of the r.h.s. of (\ref{disper}), expanding in powers of $|u_i|^2 \ll 1$ and keeping only the lowest non-trivial order, we obtain, after taking the average $\ll \dots \gg$, denoting wither higher-genus quantum fluctuations (\ref{urecflct}) for the case of a single quantum-fluctuating D-particle, or a statistical average for the case of populations of $D$-particles
affecting the $\phi$-meson decay:
\begin{equation}
   \ll \omega \gg \simeq \sqrt{k^2 + m^2} + \frac{m^2 \ll|u_i|^2 \gg}{2\sqrt{k^2 + m^2}} + \dots =
  \sqrt{k^2 + m^2} + \zeta^2 m^2 \frac{k^2}{2\sqrt{k^2 + m^2}}  + \dots
  \end{equation}
For the case of a single D-particle being present in the decay of $\phi$-meson, the parameter $\zeta \sim {\cal O}(1)$ (c.f. (\ref{urecflct}) and related discussion in that section).
In the spirit of ref.~\cite{bernabeu}, we treat the interaction term
\begin{equation}\label{interaction}
\widehat{H}_I \equiv \widehat{m}^2 \zeta^2 \frac{\widehat{k}^2}{2\sqrt{\widehat{k}^2 + \widehat{m}^2}}
\end{equation}
as an operator generating a \emph{quantum Hamiltonian perturbation} in the framework of non-degenerate
perturbation theory.
This would give the \textquotedblleft
gravitationally-dressed\textquotedblright\ initial entangled meson states,
immediately after the $\phi$ decay. The result is:
\begin{eqnarray}
&&  \left\vert {k,\uparrow}\right\rangle _{QG}^{\left(  1\right)  }\left\vert
{-k,\downarrow}\right\rangle _{QG}^{\left(  2\right)  }-\left\vert
{k,\downarrow}\right\rangle _{QG}^{\left(  1\right)  }\left\vert {-k,\uparrow
}\right\rangle _{QG}^{\left(  2\right)  }=\left\vert {k,\uparrow}\right\rangle
^{\left(  1\right)  }\left\vert {-k,\downarrow}\right\rangle ^{\left(
2\right)  }-\left\vert {k,\downarrow}\right\rangle ^{\left(  1\right)
}\left\vert {-k,\uparrow}\right\rangle ^{\left(  2\right)  }\nonumber\\
&&  +\left\vert {k,\downarrow}\right\rangle ^{\left(  1\right)  }\left\vert
{-k,\downarrow}\right\rangle ^{\left(  2\right)  }\left(  {\beta^{\left(
1\right)  }-\beta^{\left(  2\right)  }}\right)  +\left\vert {k,\uparrow
}\right\rangle ^{\left(  1\right)  }\left\vert {-k,\uparrow}\right\rangle
^{\left(  2\right)  }\left(  {\alpha^{\left(  2\right)  }-\alpha^{\left(
1\right)  }}\right)  \nonumber\\
&&  +\beta^{\left(  1\right)  }\alpha^{\left(  2\right)  }\left\vert
{k,\downarrow}\right\rangle ^{\left(  1\right)  }\left\vert {-k,\uparrow
}\right\rangle ^{\left(  2\right)  }-\alpha^{\left(  1\right)  }\beta^{\left(
2\right)  }\left\vert {k,\uparrow}\right\rangle ^{\left(  1\right)
}\left\vert {-k,\downarrow}\right\rangle ^{\left(  2\right)  }\label{entangl}%
\end{eqnarray}
where
\begin{equation}
\alpha^{\left(  i\right)  }=\frac{^{\left(  i\right)  }\left\langle
\uparrow,k^{\left(  i\right)  }\right\vert \widehat{H_{I}}\left\vert
k^{\left(  i\right)  },\downarrow\right\rangle ^{\left(  i\right)  }}%
{E_{2}-E_{1}}~,\quad\beta^{\left(  i\right)  }=\frac{^{\left(  i\right)
}\left\langle \downarrow,k^{\left(  i\right)  }\right\vert \widehat{H_{I}%
}\left\vert k^{\left(  i\right)  },\uparrow\right\rangle ^{\left(  i\right)
}}{E_{1}-E_{2}}~,~\quad i=1,2\label{qgpert2}%
\end{equation}and the index $(i)$ runs over meson species (\textquotedblleft
flavours\textquotedblright) ($1\rightarrow K_{L},~2\rightarrow K_{S}$). The
reader should notice that the terms proportional to $\left(  {\alpha^{\left(
2\right)  }-\alpha^{\left(  1\right)  }}\right)  $ and $\left(  {\beta
^{\left(  1\right)  }-\beta^{\left(  2\right)  }}\right)  $ in (\ref{entangl})
generate $\omega$-like effects, if the coefficients are non zero.

In \cite{bernabeu} we have discussed models within the context of D-particle foam, where such terms are generated in the initial state, as a result of space-time metric distortions that have off diagonal $g_{0i}$ components proportional to the D-particle recoil velocity $u_i$.
As explained in detail in \cite{bernabeu}, where we refer the interested reader for details, $\omega$-like terms are present in the initial entangled states of Kaons after the $\phi$-decay, provided the interactions of a Kaon state with a D-particle change the mass eigenstate (``flavour'' changing interaction), in other words the re-emitted open string after capture in fig.~\ref{fig:recoil} is characterised by a different mass than the incident Kaon, while momentum is conserved on average. In such a case, the gravitational dressing
(\ref{entangl}) can be achieved~\cite{bernabeu} by flavour-changing perturbations of the form:
{\small \begin{equation}\label{pertvec} {\widehat H}_I^{\rm other~models} = -\left(  { r_{1} \sigma_{1} + r_{2} \sigma_{2}}
\right) \widehat{k}~, \quad \ll r_i \gg =0~, i=1,2,3 \, \ll r_i r_j \gg = \delta_{ij} \sigma_0~.
\end{equation}}
where $\ll \dots \gg $ denote appropriate averages over stochastic and population effects of the foam, as usual.  The reader should notice the vector nature in momentum space of this perturbation, which as explained in detail in \cite{bernabeu}, is a consequence of the off-diagonal metric elements $g_{0i} \sim u_i $ that affect the probe's dispersion relation appropriately, leading -- after perturbative expansion in powers of $u_i$, as above -- to the form (\ref{pertvec}).

We next remark that on averaging the matter-probe density matrix over the random variables
$r_{i}$, which are treated as independent variables between the two meson
particles of the initial state (\ref{entangl}), we observe that only terms of
order $|{\omega}|^{2}$ will survive, with the order of $|{\omega}|^{2}$ being
\begin{eqnarray}
&& |{\omega}|^{2}=\widetilde{\sum}_{(1),(2)}\left(  \mathcal{O}\left(
\frac{1}{(E_{1}-E_{2})^2}(\langle\downarrow,k|H_{I}^{\rm other_models}|k,\uparrow\rangle
)^{2}\right)  \right)  = \nonumber \\
&& \widetilde{\sum}_{(1),(2)}\left(  \mathcal{O}\left(
\frac{\sigma_{0}k^{2}}{(E_{1}-E_{2})^{2}}\right)  \right) \, \sim \, \widetilde
{\sum}_{(1),(2)}\left(  \frac{\sigma_{0}k^{2}}{(m_{1}-m_{2})^{2}}\right)
\label{omegaorder}%
\end{eqnarray}
for the physically interesting case of non-relativistic Kaons in $\phi$
factories, in which the momenta are of order of the rest energies. The
notation $\widetilde{\sum}_{(1),(2)}\left(  \dots\right)$ above indicates
that one considers the sum of the variances $\sigma_{0}$ over the
two meson states $1$, $2$ as defined above. The latter can be estimated in a similar way as in
our case here, (\ref{recvelquantuflct}), leading to the following order of magnitude estimate of the $\omega$-effects in the initial state~\cite{tsallissarkar}:
\begin{equation}
|\omega|^{2}\sim \xi^2 g_{s}^{2}\frac{\left(  m_{1}^{2} + m_{2}^{2}\right)  }%
{M_{s}^{2}}\frac{k^{2}}{(m_{1}-m_{2})^{2}}~,\label{finalomega}%
\end{equation}where the factor $\xi^2$ takes proper account of statistical (over populations of D-particles) effects,
that might be present during the initial decay of the $\phi$-meson. As already mentioned, for the case of a single D-particle present during the $\phi$-meson decay, this factor is of order $\xi ={\cal O}(1)$, if substructure of the mesons is ignored when quantum gravitational interactions are considered.
In realistic situations, however, where the strong interaction substructure of the Kaons is taken into account, such effects are also absorbed (in a sort of mean-field way) into this parameter, which thus may no longer be of order one, even for a single fluctuating D-particle. In fact, there might be a strong-interaction suppression of the effects due to the D-particle interactions with the (electrically neutral) gluon constituents of the mesons, in which case $\xi \ll 1$. At present, such detailed calculations have not been performed.

The result (\ref{finalomega}), implies, for neutral Kaons in a $\phi$ factory
($m_{L} - m_{S} \sim 3.48 \times10^{-15}~\mathrm{GeV}$),
the following estimate~\cite{tsallissarkar}
\begin{equation}\label{omegaf}
|\omega| =\xi {\cal O}\left( 10^{-5}\right)~,
\end{equation}
As we shall discuss in subsection \ref{sec:exp}, such effects can in principle be falsifiable in the next generation facilities, provided $\xi $ is of order ${\cal O}(1)$.
Thus, we see that the near degeneracy of the two mass-eigenstates of the neutral mesons, $(m_1 - m_2)/m_1 \ll 1$, provides the appropriate \emph{magnifying} effects of an otherwise tiny quantum-gravity effect, suppressed by the square of the quantum-gravity mass scale, here the mass $M_s/g_s$ of the D-particle defect in the foam.
A similar r\^ole was played in the decoherence-induced $\omega$-effect (\ref{gammadec}) by the near-zero width difference $\Delta \Gamma \sim 10^{-15} $~GeV of the two states.

In the model we are considering here, however, where the induced metric is diagonal (\ref{opsmetric}), the form of the interaction (\ref{interaction}) does not change, even if we allow for ``flavour'' changing-interactions between the D-particles and the matter strings.
Indeed, even if we assume the correspondence (\ref{mutliflavour}), according to which $u_i \to r_\mu \sigma_\mu $, where $\sigma_\mu$ an appropriate basis of $2 \times 2 $ matrices, including the Pauli ones, we observe that,
for the stochastically fluctuating Gaussian case we assume in this work, in which
$\ll r_\mu \gg =0, ~\ll r_\mu r_\nu \gg = \delta_{\mu\nu} \zeta^2 $,
 the square $\ll u_i^2 \gg$,
always remains proportional to the identity in flavour space.
Hence, in Eq.~(\ref{initialomega}), the coefficients $\alpha^{(i)}=\beta^{(i)} = 0$, and thus no $\omega$-effects appear in the initial state  as a result of this gravitational dressing in the model considered here.

 Nevertheless, by allowing other more general interactions, for instance (\ref{pertvec}) based on off-diagonal induced metrics~\cite{bernabeu,tsallissarkar}, one can easily obtain $\omega$-like terms in the initial state. Hence one should always keep an open mind about this issue, especially when performs generic phenomenological searches of such quantum-gravity effects.
If an $\omega$-term is present in the initial state (\ref{initialomega}), then a decoherent Lindblad evolution evolution, parametrized by (\ref{ehnsdec}), (\ref{cons})  will generate terms of the form (\ref{gammadec}) but with modified coefficients~\cite{bmpw}:
\begin{equation}
\left(|\omega|^2 -\frac{2\gamma}{\Delta \Gamma}\right) \rho_S \otimes \rho_S~, \quad  \left(|\omega|^2 + \frac{2\gamma}{\Delta \Gamma} \right)\rho_L \otimes \rho_L~
\label{gammadec2}
\end{equation}
A detailed analysis of various physically interesting observables in a $\phi$-factory, including
identical final states, has been performed in \cite{bmpw}, where we refer the reader for details on the form and the magnitude of the $\omega$-like effects.

\subsubsection{$\omega$-Effect Observables in $\phi$-factories\label{sec:omegaobs}}

To construct the appropriate observable for the possible detection
of $\omega$-like effects, we consider the $\phi$-decay amplitude
depicted in Fig.~\ref{epr}, where one of the kaon products decays to
the  final state $X$ at $t_1$ and the other to the final state $Y$
at time $t_2$. We take $t=0$ as the moment of the $\phi$-meson
decay.

The relevant amplitudes read:
\begin{eqnarray*}
A(X,Y) = \langle X|K_S\rangle \langle Y|K_S \rangle \, {\cal N}
\,\left( A_1  +  A_2 \right)~, \nonumber
\end{eqnarray*}
with \begin{eqnarray*}
 A_1  &=& e^{-i(\lambda_L+\lambda_S)t/2}
[\eta_X  e^{-i \Delta\lambda \Delta t/2}
-\eta_Y  e^{i \Delta\lambda \Delta t/2}]\nonumber \\
A_2  &=&  \omega [ e^{-i \lambda_S t} - \eta_X \eta_Y e^{-i
\lambda_L t}] \nonumber
\end{eqnarray*}
denoting the CPT-allowed and CPT-violating parameters respectively,
and $\eta_X = \langle X|K_L\rangle/\langle X|K_S\rangle$ and $\eta_Y
=\langle Y|K_L\rangle/\langle Y|K_S\rangle$. In the above formulae, $t$
is the sum of the decay times $t_1, t_2$ and $\Delta t $ is their
difference (assumed positive).

\begin{figure}[htb]
\begin{center}
  \includegraphics[width=0.5\textwidth]{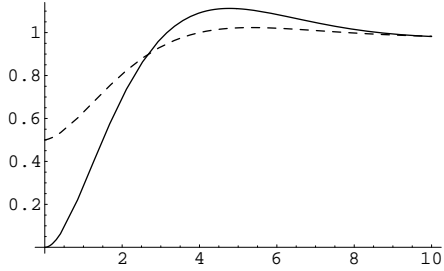}
\end{center}
\caption{A characteristic case of the intensity $I(\Delta t)$, with
$|\omega|=0$ (solid line)  vs  $I(\Delta t)$ (dashed line) with
$|\omega|=|\eta_{+-}|$, $\Omega = \phi_{+-} - 0.16\pi$, for
definiteness~\cite{bmp}.} \label{intensomega}\end{figure}

The ``intensity'' $I(\Delta t)$ is the
desired \emph{observable} for a detection of the $\omega$-effect,
\begin{eqnarray} \label{omegevoldet} I (\Delta t) \equiv \frac{1}{2} \int_{\Delta
t}^\infty dt\, |A(X,Y)|^2~.
\end{eqnarray}
depending only on $\Delta t$.

Its time profile reads~\cite{bmp}:
\begin{eqnarray}\label{omegevoldet2} &&
I (\Delta t) \equiv \frac{1}{2} \int_{|\Delta t|}^\infty dt\,
|A(\pi^+\pi^-,\pi^+\pi^-)|^2  = \nonumber
\\ && |\langle\pi^{+}\pi^{-}|K_S\rangle|^4 |{\cal N}|^2 |\eta_{+-}|^2
\bigg[ I_1  + I_2  +  I_{12} \bigg]~,
\end{eqnarray}
where
\begin{eqnarray}\label{omegevoldet3} && I_1 (\Delta t) =
\frac{e^{-\Gamma_S \Delta t} + e^{-\Gamma_L \Delta t} - 2
e^{-(\Gamma_S+\Gamma_L) \Delta t/2} \cos(\Delta m \Delta t)}
{\Gamma_L+\Gamma_S}
\nonumber \\
&& I_2 (\Delta t) =  \frac{|\omega|^2 }{|\eta_{+-}|^2}
\frac{e^{-\Gamma_S \Delta t} }{2 \Gamma_S}
\nonumber \\
&& I_{12} (\Delta t) = - \frac{4}{4 (\Delta m)^2 + (3 \Gamma_S +
\Gamma_L)^2}  \frac{|\omega|}{|\eta_{+-}|} \times
\nonumber \\
&&\bigg[ 2 \Delta m \bigg( e^{-\Gamma_S \Delta t} \sin(\phi_{+-}-
\Omega) - \nonumber \\
&&  e^{-(\Gamma_S+\Gamma_L) \Delta t/2} \sin(\phi_{+-}- \Omega
+\Delta m \Delta t)\bigg)
\nonumber \\
&&  - (3 \Gamma_S + \Gamma_L) \bigg(e^{-\Gamma_S \Delta t}
\cos(\phi_{+-}- \Omega) - \nonumber \\
&& e^{-(\Gamma_S+\Gamma_L) \Delta t/2} \cos(\phi_{+-}- \Omega
+\Delta m \Delta t)\bigg)\bigg]~,
\end{eqnarray}
with $\Delta m = m_S - m_L$ and $\eta_{+-}= |\eta_{+-}|
e^{i\phi_{+-}}$ in the usual notation~\cite{fide}.
A typical case for the relevant intensities, indicating clearly the
novel CPTV $\omega$-effects, is depicted in Fig.~\ref{intensomega}.

As seen from (\ref{omegevoldet3}), the novel $\omega$-effect appears in
the combination $\frac{|\omega|}{|\eta_{+-}|}$, thereby implying
that the decay channel to $\pi^+\pi^-$ is particularly sensitive to
the $\omega$ effect~\cite{bmp}, due to the enhancement by
$1/|\eta_{+-}| \sim 10^{3}$, implying sensitivities up to
$|\omega|\sim 10^{-6}$ in $\phi$ factories. The physical reason for
this enhancement is that $\omega$ enters through $K_SK_S$ as opposed to
$K_LK_S$ terms, and the $K_L \to \pi^+\pi^-$ decay is CP-violating.
Although above we considered $\omega$-like terms in the initial state of the entangled Kaons after $\phi$-decay, and a unitary hamiltonian evolution for simplicity, qualitatively similar results
pertain to our case of D-particle foam, where $\omega$-like effects are generated by Lindblad decoherence (\ref{gammadec}). For a complete list of observables in $\phi$-factories, and detailed analysis of testing
$\omega$-like effects, including those generated by decoherent evolution, in $\phi$-factories we refer the reader to ref.~\cite{bmpw}.

\subsubsection{Experimental Bounds on $\omega$-like and Decoherence Effects\label{sec:exp}}

Experimentally, the situation concerning the most recent bounds on $\gamma$ and $|\omega|$ parameters can be summarised as follows:
the KLOE experiment at DA$\Phi$NE
has released the latest measurement of the $\omega$
parameter~\cite{dafne}:
\begin{eqnarray}\label{kloe} &&
{\rm Re}(\omega) = \left(
-1.6^{+3.0}_{-2.1} \pm 0.4\right)\times 10^{-4}~, \quad
{\rm Im}(\omega) = \left( -1.7^{+3.3}_{-3.0} \pm 1.2\right)\times
10^{-4}~, \nonumber \\ && |\omega | < ~1.0 \times 10^{-3}~~{\rm at~95~\%~C.L.}\end{eqnarray}
One can constrain the $\omega$
parameter significantly in
upgraded facilities. For instance, there are the following
perspectives for KLOE-2 at (the upgraded)
DA$\Phi$NE-2~\cite{dafne}:
\begin{equation}\label{prospects}
{\rm Re}(\omega),~ {\rm Im}(\omega) \longrightarrow 2 \times 10^{-5}~.
\end{equation}
Thus we see that sjuch searches can indeed falsify some models of D-particle foam where $\omega$-effects in the initial state arise as a result of foam effects on the decay of the $\phi$-meson, for which the estimate (\ref{finalomega}), (\ref{omegaf}) is valid, provided $\xi = {\cal O}(1)$.

On the other hand, the Lindblad decoherence parameter $\gamma$, in completely positive parameterizations (\ref{ehnsdec}), (\ref{cons}),
can be constrained with the highest-possible sensitivity at present in $\phi$-factories at DA$\Phi$NE, since
the KLOE experiment has the greatest
sensitivity to this parameter $\gamma$. The latest KLOE
measurement for $\gamma$, as reported in ref.~\cite{dafne},
yields
\begin{equation}
\gamma_{\rm
KLOE} = (0.7^{+1.2}_{-1.2} \pm 0.3) \times 10^{-21}~{\rm GeV}~,
\label{gammakloe}
\end{equation}
\emph{i.e.}
$\gamma < 7 \times 10^{-22}~{\rm GeV}$, competitive with the
corresponding CPLEAR bound~\cite{cplear}. It is
expected that this bound could be improved by an order of magnitude
in the upgraded facilities KLOE-2 at
DA$\Phi$NE-2~\cite{dafne}, where one expects
\begin{equation}\label{uprgade}
\gamma_{\rm
upgrade} \to \pm 0.2 \times 10^{-21} ~{\rm GeV}~.
\end{equation}
In our decoherence-induced D-particle model, where the theoretical estimate (\ref{dampinglindgamma}) is valid,
the above bounds imply
\begin{equation}
 \frac{M_s}{g_s^2 \zeta^2 } \, >\, 5 \times 10^{21} \, {\rm GeV}~.
 \end{equation}
or, equivalently, $\zeta^2 < 5 \times 10^{-4}$~, making the natural assumption for the value $M_s/g_s \sim 10^{19}$~GeV (Planck scale) of the mass of the D-particles. Thus, if the microscopic models are characterised by parameters $(\xi, \zeta) \sim {\cal O}(1)$, then, they can be falsified in these upgraded neutral-meson facilities.

However,  as already mentioned, the limits for the decoherence parameter $\gamma$ coming from neutrino oscillation experiments~\cite{neutrino}, with the energy dependence (\ref{dampinglindgamma})~\cite{neutrino}, are stronger by several order of magnitudes. In this sense, next generation neutral Kaon facilities  might not have the sensitivity to falsify such models, provided of course that the decoherent effects act \emph{universally} among neutral kaons and neutrinos.

\section{Conclusions and Outlook \label{sec:4}}

This work examined the r\^ole of quantum string fluctuations (which can give
rise to a non-commutative space-time geometry at string scales) on the velocity distribution of
D-particles, within a specific kind of foam in string theory. There is induced decoherence
for low-energy string matter propagating in this stochastically fluctuating space time background, and as a result a fundamental arrow of time, manifested through an ill-defined CPT generator. This implies intrinsic
CPT Violation of a rather unconventional kind.
It should be stressed that the CPT Violation does not arise from the space-time non commutativity \emph{per se}, but it is due to the deviation of the pertinent $\sigma$-model, describing first quantised strings in such a background, from the world-sheet conformal point, which in turn induces quantum decoherence in target space.

Our Gaussian modeling of the recoil velocity is found to be
robust to these fluctuations. In this way we have managed to give a rather rigorous estimation (modulo
strong interaction effects) of decoherence-induced $\omega$-like effects, associated with modifications of EPR correlations of entangled states of neutral mesons. For certain simplistic and overoptimistic models of D-particle foam such effects are of a magnitude that might make these models falsifiable at the next generation meson-factory facilities, such as an upgrade of DA$\Phi$NE. However, if one accepts the universality of quantum gravity (at least on electrically neutral probes) it appears that limits coming from neutrino experiments indicate much stronger suppression of decoherence effects, which indeed would imply a much more dilute population of space-time defects. If this is the case, then potential detection of the decoherence-induced $\omega$-like terms in entangled states of future meson factories may not be feasible in the foreseeable future.

Admittedly, our approach in this paper is based on bosonic string theory
which is not the most relevant phenomenologically. World-sheet supersymmetric
strings do not lead to a closed-form resummation of the leading divergencies of the pinched surfaces, since the latter cancel out~\cite{szabo2,mav2}, and the remaining terms are hard to
cast in a closed form.
However, despite this apparent technical difficulty,  the general
conclusions drawn from the current work, as far as \emph{fuzzyness} of the target space time is concerned,
as well as its Finsler-type, due to the dependence of the metric distortions on the momentum transfer during the interaction of the D-particle with the open-string matter,
are likely to be robust, since they depend on the form of vertex
operators for (recoil) zero modes of D-particles (provided of course the theories are
restricted to those admitting D-particles).

As we have discussed above, the interaction of D-particles with matter leads to local distortions of the neighboring space-time of Finsler type, depending on both the string coupling $g_s$, through the D-particle masses $M_s/g_s$, and the recoil velocity of the D-particle (i.e. the momentum transfer
of the string matter) $u_i$, which stochastically fluctuates upon summing up higher-genus
world-sheet topologies. In view of our discussion in this work, both these quantities can be ``fuzzy'', leading to stochastic fluctuations on the space-time metric, on which
string matter lives. These fluctuations are up and above any statistical fluctuations in populations of D-particles that characterise the D-particle foam models.
This is an important aspect of the formalism discussed here, since it may have a profound influence on the dark matter (and dark energy) distributions in a Universe with D-particle foam, which may have phenomenological consequences, as far as constraints on, say, supersymmetric particle physics models are concerned.

The presence of such fluctuations affect important cosmological
quantities that are directly relevant to the string-Universe energy budget, such as thermal supersymmetric dark matter relic densities, through appropriate modifications of the
relevant thermodynamic equations. Hence, the relevant astro-particle physics constraints on supersymmetric models~\cite{lmn} are also modified. However, the issue as to whether the fuzzyness of the D-particle foam space-time can lead to observable signatures in Cosmology or astro-particle physics in general, remains to be seen. We hope to be able to report in a more detailed form on these phenomenological issues in the near future.

\section*{Acknowledgements}

I thank E. Milotti and the other organisers of the \emph{SPIN-STAT 2008} Conference  (Trieste, October 21-25 2008) for the invitation to this very interesting event, where results from the current work have been presented. This work is partially supported by the European Union through the FP6 Marie
Curie Research and Training Network \emph{UniverseNet} (MRTN-CT-2006-035863).

\end{document}